\documentclass[twocolumn]{aastex631}

\usepackage{amsmath}
\usepackage[graphicx]{realboxes}

\newcommand\xISM{\texttt{ExoTiC-ISM}\xspace}
\newcommand\xLD{\texttt{ExoTiC-LD}\xspace}
\newcommand\GRISMCONF{\texttt{GRISMCONF}\xspace}
\newcommand\Hazelnut{\texttt{hazelnut}\xspace}
\newcommand\lluvia{\texttt{lluvia}\xspace}

\usepackage{latexsym, graphicx, amssymb, longtable, epsf, xspace}
\usepackage[version=4]{mhchem}

\begin{document}


\title{The HUSTLE Program: The UV to Near-Infrared HST WFC3/UVIS G280 \\ Transmission Spectrum of WASP-127b}

\author[0000-0002-4945-1860]{V. A. Boehm}
\affiliation{Department of Astronomy, Cornell University, 122 Sciences Drive, Ithaca, NY 14853, USA}

\author[0000-0002-8507-1304]{N. K. Lewis}
\affiliation{Department of Astronomy, Cornell University, 122 Sciences Drive, Ithaca, NY 14853, USA}

\author[0000-0001-9665-5260]{C. E. Fairman}
\affiliation{School of Physics, University of Bristol, H.H. Wills Physics Laboratory, Tyndall Avenue, Bristol BS8 1TL, UK}

\author[0000-0002-6721-3284]{S. E. Moran}
\affiliation{Department of Planetary Sciences and Lunar and Planetary Laboratory, University of Arizona, Tuscon, AZ 85721, USA}

\author[0000-0001-5097-9251]{C. Gascón}
\affiliation{Center for Astrophysics ${\rm \mid}$ Harvard {\rm \&} Smithsonian, 60 Garden St, Cambridge, MA 02138, USA}
\affiliation{Institut d'Estudis Espacials de Catalunya (IEEC), 08860 Castelldefels (Barcelona), Spain}

\author[0000-0003-4328-3867]{H. R. Wakeford}
\affiliation{School of Physics, University of Bristol, H.H. Wills Physics Laboratory, Tyndall Avenue, Bristol BS8 1TL, UK}

\author[0000-0003-4157-832X]{M. K. Alam}
\affiliation{Space Telescope Science Institute, 3700 San Martin Drive, Baltimore, MD 21218, USA}

\author[0000-0001-8703-7751]{L. Alderson} 
\affiliation{School of Physics, University of Bristol, HH Wills Physics Laboratory, Tyndall Avenue, Bristol BS8 1TL, UK}

\author[0000-0003-3726-5419]{J. Barstow} 
\affiliation{School of Physical Sciences, The Open University, Walton Hall, Milton Keynes MK7 6AA, UK}

\author[0000-0003-1240-6844]{N. E. Batalha} 
\affiliation{NASA Ames Research Center, Moffett Field, CA 94035, USA}

\author[0000-0001-5878-618X]{D. Grant} 
\affiliation{HH Wills Physics Laboratory, University of Bristol, Tyndall Avenue, Bristol, BS8 1TL, UK}

\author[0000-0003-3204-8183]{M. L\'opez-Morales} 
\affiliation{Center for Astrophysics ${\rm \mid}$ Harvard {\rm \&} Smithsonian, 60 Garden St, Cambridge, MA 02138, USA}

\author[0000-0003-4816-3469]{R. J. MacDonald}
\altaffiliation{NHFP Sagan Fellow}
\affiliation{Department of Astronomy, University of Michigan, 1085 S. University Ave., Ann Arbor, MI 48109, USA}

\author[0000-0002-5251-2943]{Mark S. Marley}\affiliation{Department of Planetary Sciences and Lunar and Planetary Laboratory, University of Arizona, Tuscon, AZ 85721, USA}

\author[0000-0003-3290-6758]{K. Ohno} 
\affiliation{Department of Astronomy and Astrophysics, University of California, Santa Cruz, Santa Cruz, CA, USA}










\begin{abstract}

Ultraviolet wavelengths offer unique insights into aerosols in exoplanetary atmospheres. However, only a handful of exoplanets have been observed in the ultraviolet to date. Here, we present the ultraviolet-visible transmission spectrum of the inflated hot Jupiter WASP-127b. We observed one transit of WASP-127b with WFC3/UVIS G280 as part of the Hubble Ultraviolet-optical Survey of Transiting Legacy Exoplanets (HUSTLE), obtaining a transmission spectrum from 200--800~nm. Our reductions yielded a broad-band transit depth precision of 91~ppm and a median precision of 240~ppm across 59 spectral channels. 
Our observations are suggestive of a high-altitude cloud layer with forward modeling showing they are composed of sub-micron particles and retrievals indicating a high opacity patchy cloud. 
While our UVIS/G280 data only offers weak evidence for Na, adding archival HST WFC3/IR and STIS observations raises the overall Na detection significance to $4.1\sigma$. Our work demonstrates the capabilities of HST WFC3/UVIS G280 observations to probe the aerosols and atmospheric composition of transiting hot Jupiters with comparable precision to HST STIS.

\end{abstract}



\section{Introduction} \label{sec:intro}
The past decade of research into the atmospheres of transiting hot Jupiters has made thorough use of the \textit{Hubble Space Telescope} (HST) and its Wide Field Camera 3 (WFC3) infrared (IR) channel and the Space Telescope Imaging Spectrograph (STIS). In particular, studies regularly made use of the STIS G430L and G750L gratings (300~nm to 1050~nm), and the WFC3/IR G141 grism (1100~nm to 1600~nm). Observations in these modes were sometimes supplemented with the WFC3/IR G102 grism (800~nm to 1100~nm) to bridge the gap between STIS and WFC3/IR (see e.g., \citealt{STIS2013}, \citealt{STIS2015}, \citealt{Wakeford2018}, \citealt{Spake2021}). Additionally, the high-resolution echelle gratings (e.g., STIS E230M, see \citealt{Sing2019}) were also used to probe escaping atmospheres. By combining multiple transit observations, each taken through a different mode, a detailed transmission spectrum spanning 300~nm to 1700~nm could be built up. However, as this observing strategy requires multiple transits, it can be subject to variability in instrumental systematics and stellar activity across visits.

The commonly used HST observing modes barely measure ultraviolet (UV) spectra of exoplanet atmospheres (100--400~nm). The near- (300--400~nm) and mid-UV (200--300~nm) spectra of hot Jupiters are an underutilized probe of atmospheric aerosols, as short UV wavelengths show particularly strong signatures of Rayleigh scattering caused by nm- to \textmu m-scale particles \citep[e.g.,][]{Ohno2020}. Furthermore, the UV wavelengths cover a range of strongly absorbing gasses and molecules such as silicate oxide vapor \citep[e.g.,][]{Lothringer2022} or mineral  condensates \citep[e.g.,][]{WakefordSing2015}, which are important for overall atmospheric chemistry. 

The HST WFC3/UVIS G280 grism is an ultraviolet-visible (UVIS) grism that went largely unused by the exoplanetary science community until recently. \cite{Wakeford2020} demonstrated WFC3/UVIS G280's capabilities for exoplanet characterization via a proof-of-concept study on the hot Jupiter HAT-P-41b. Several unusual features of UVIS/G280 make it challenging to work with, in particular its strongly curving traces and its higher orders that overlap the lower orders and introduce (typically minor) spectral self-contamination. Nevertheless, \cite{Wakeford2020} and additional studies (e.g., \citealt{Lothringer2022}) have proven that, despite these quirks, UVIS/G280 can produce reliable observations in the near- and mid-UV that make it worth the challenge. The spectral range of the UVIS/G280 grism (200--800~nm), which can be covered in just one transit observation, is comparable to the combined spectral range of both the STIS G430L and G750L instruments (300--1050~nm), which require at least two transit observations to acquire. UVIS/G280 also has superior throughput from 200--300~nm when compared to other HST modes, enabling us to constrain exoplanet cloud properties and atmospheric escape processes more robustly than has been done with other instruments. The proof-of-concept study presented in \cite{Wakeford2020} set the groundwork for the \textit{Hubble} Ultraviolet-optical Survey of Transiting Legacy Exoplanets (HUSTLE) program, an ongoing HST treasury program observing 12 hot Jupiters using WFC3/UVIS G280 to probe transitions in atmospheric chemistry and structure across equilibrium temperatures spanning 960--2640~K \citep{HUSTLE2022}.

As part of HUSTLE, we obtained observations of the hot Jupiter WASP-127b through UVIS/G280. WASP-127b is a $T_{eq}=1400\pm24$~K hot Jupiter with a highly inflated radius of $R_p = 1.37\pm0.04R_J$, despite being less massive than Saturn at $M_p = 0.18\pm0.02M_J$ \citep{Lam2017}. The scale height of its atmosphere is predicted to be in excess of $2000$~km \citep{Spake2021}, one of the largest yet discovered. WASP-127b orbits a $V\sim10.2$, photometrically-quiet G dwarf star \citep{Lam2017}. The host star's brightness and low activity level, combined with the large expected atmospheric scale height of the planet, make WASP-127b an ideal target for UVIS transmission spectroscopy. Furthermore, several studies posit that at $T_{eq}\sim1400$~K, hot Jupiters and brown dwarfs may undergo a transition in aerosol structure, wherein silicate clouds begin to subside and manganese sulfide clouds (MnS) begin to condense in the upper atmosphere \citep[e.g.,][]{Marley2012,Parm2016,Wakeford2017}. In contrast, microphysics studies \citep[e.g.,][]{Gao2020} suggest that energy barriers will inhibit the nucleation of MnS condensates so that MnS clouds should not be expected to be dominant on WASP-127b. Studies of this world with UVIS/G280 can constrain the nature of the clouds in its atmosphere, which will serve as a test of whether there is a transition from silicate clouds to MnS clouds near 1400\,K.

Previous studies have targeted WASP-127b using data from both ground-based instruments --- such as the Nordic Optical Telescope \citep{Palle2017}, the Gran Telescopio Canarias \citep{Chen2018}, HARPS \citep{Zak2019}, and ESPRESSO \citep{Allart2020} --- and from space-based observations with HST (300--1700~nm), \textit{Spitzer} photometry at 3.6 and 4.5\,\textmu m, and TESS broadband photometry in the red optical \citep{Skaf2020,Spake2021}. These efforts characterized the spectrum of WASP-127b down to 300~nm and found a rise in transit depth at wavelengths short of 560~nm, but all studies disagreed on the steepness of this rise. \citet{Palle2017} and \citet{Chen2018} found a super-Rayleigh slope.
To our current knowledge, the super-Rayleigh slope could be explained only by photochemical hazes with vigorous vertical mixing \citep{Kawashima&Ikoma19,Ohno2020}, optical and/or NUV absorbers such as SH radical \citep{Evans+18} and sulfide clouds \citep[e.g., MnS, see][]{Pinhas2017}, or uncorrected contamination by starspots \citep[e.g.,][]{McCullough+14}, although the last explanation is somewhat disfavored for WASP-127b by the star's reported quietness \citep{Lam2017}. In contrast, \cite{Spake2021} observed a shallower, sub-Rayleigh scattering slope.
If their finding is true, it could considerably alter the interpretation of the spectrum, as there are several other possibilities to explain the sub-Rayleigh slope, including mineral clouds without any requirement for particular cloud compositions \citep[e.g.,][]{Pinhas2017,Powell+19,Ormel&Min19,Ohno+20_fluffy} as well as photochemical hazes without vigorous atmospheric mixing, especially for soot-like haze compositions \citep[e.g.][]{Lavvas&Koskinen17,Ohno2020,Steinrueck+23}.

Regarding the atmospheric compositions of WASP-127b, probes into the near-IR spectrum through WFC3/IR and \textit{Spitzer} have indicated super-solar abundances of H$_2$O and the possible presence of either CO$_2$ or CO \citep{Skaf2020,Spake2021}. This was followed up with ground-based observations with SPIRou from 900--2500~nm, which found CO at sub-solar abundances and an H$_2$O abundance closer to solar \citep{Boucher2023}. Ground-based spectroscopy spanning 400--1000~nm suggested the presence of alkali metals in the atmosphere of WASP-127b, such as sodium, lithium, and potassium \citep{Palle2017, Chen2018, Zak2019, Allart2020}; however, space-based spectroscopic studies that confirmed the presence of sodium found no definitive evidence of lithium or potassium \citep{Spake2021}, which can be contaminated by telluric absorption from the ground. 

Our study expands on this narrative by using HST's WFC3/UVIS G280 observing mode to resolve the spectrum of WASP-127b down to 200~nm with increased near-UV precision. In doing so, we aim to place stronger constraints on the spectrum short of 560~nm, which will better reveal the nature of the aerosols responsible for the rise in transit depth at these wavelengths. 
Furthermore, the spectral range of HST's WFC3/UVIS G280 
will allow for the reexamination of prior claims of sodium, lithium, and potassium in the atmosphere of WASP-127b. 
We jointly analyze our new WFC3/UVIS G280 spectrum alongside archival data from HST's STIS, WFC3/IR, and \textit{Spitzer} photometry to probe WASP-127b's atmosphere from the mid-UV to the near-IR. 

Our study is structured as follows. In Section~\ref{sec:data}, we present our observations, our data reduction strategies, and our methods for extracting transit light curves from these data. Section~\ref{sec:analysis} presents the analysis of the extracted light curves, including fitting both broad-band and spectroscopic light curves to extract the system parameters and transit spectrum. In Section~\ref{sec:retrieval}, we present forward modeling and retrieval interpretations of these data, and, in Section~\ref{sec:disc}, we discuss our results and their significance in the context of previous observations of WASP-127b. Section~\ref{sec:outro} summarizes our work in the context of both HUSTLE and exoplanet studies as a whole, with suggestions for further work on this topic.

\section{Observations and Methods} \label{sec:data}

\subsection{HST WFC3/UVIS G280 Observations}
We observed one transit of the target WASP-127b with the HST WFC3/UVIS G280 grism, which provides wavelength coverage from 200~nm to 800~nm. These data were taken from 21:10:59 UTC on 2023-02-13 to 04:33:44 on 2023-02-14 as part of visit 12 of GO 17183 (PI: Hannah Wakeford). The visit spanned five HST orbits centered on mid-transit, with one orbit before and one after transit to ensure adequate pre- and post-transit baseline measurements. We collected 15, 78-second frames in the first orbit and 16 frames in every subsequent orbit for a total of 79 frames and a 49\% duty cycle. The strong curvature of UVIS/G280 traces required observations to be taken in stare mode, so no spatial scans were collected. Furthermore, because WFC3/UVIS uses a CCD detector, no non-destructive reads were available for analysis. Both the positive $m>0$ and negative $m<0$ orders of the trace were placed near the middle of chip 2 in a subarray region of dimensions 600~$\times$~2100 out of the entire 2051~$\times$~4096. Prior to UVIS/G280 measurements, a single frame was collected using the F300X UVIS filter to aid in wavelength calibration by measuring the exact position of the target in the custom subarray. 

\subsection{Data Reduction}
To verify that the results we obtained with our analysis are grounded in truth and not artifacts of our reduction process, we processed our WASP-127b UVIS/G280 observations with two independent pipelines. Performing reductions of a dataset with multiple pipelines allows one to disentangle which anomalies are intrinsic to the data and which are artifacts of the reduction process. In recent years, the importance of providing more than one reduction of a transit or eclipse observation using multiple distinct reduction pipelines has been recognized (see e.g. \citealt{ERS2023}). We present here the HST WFC3/UVIS G280 pipeline \Hazelnut and detail it and its results in Section~\ref{sec:P1}. We also present the results of a second UVIS/G280 pipeline, called \lluvia, in Section~\ref{sec:P2}.

\subsubsection{\Hazelnut} \label{sec:P1}
\begin{figure*}[ht!]
    \centering
    \includegraphics[width=0.99\textwidth]{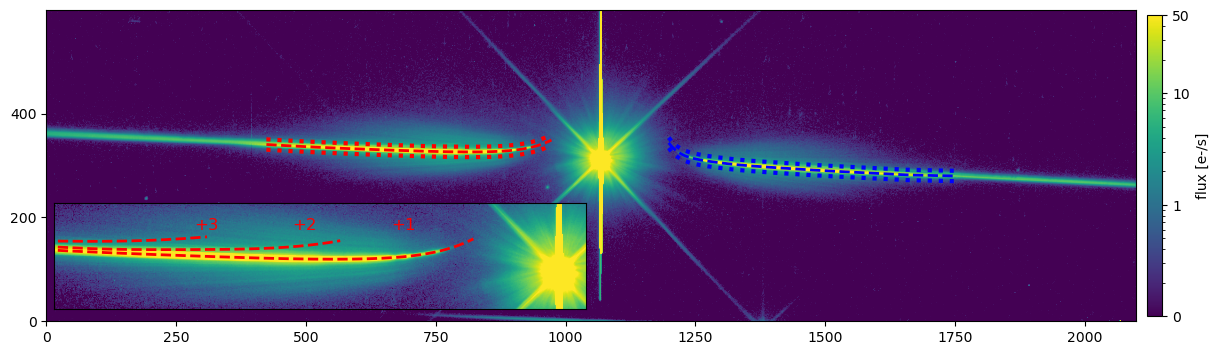}
    \caption{WASP-127 as seen through HST's WFC3/UVIS G280 grism. The saturated object seen at the center is the 0th order; the two diagonally-crossing lines are diffraction spikes and thus an artifact of the optics, while the vertical line is bloom, an artifact of severe 0th order saturation caused by electrons from very saturated pixels spilling out into vertically adjacent pixels. The gently swooping bright lines crossing the entire frame diagonally are the traces we aim to extract a spectrum from. To the left, the positive orders are dispersed, with the brightest and innermost trace being the $+1$st order. To the right, the negative orders are dispersed. Red (blue) dashed lines mark the \GRISMCONF polynomial fits and dotted lines mark the edges of the extraction aperture for the $+1$st ($-1$st) order. Background stars and their diffraction spikes/traces are faintly visible in some regions of the image; these do not affect our observations of WASP-127b. Inset: The first through third positive orders. Self-contamination at the level of 2\% or less in the $1$st order by the $2$nd order begins at about 450~nm.}
    \label{fig:frame}
\end{figure*}

Our \Hazelnut reduction followed the same procedures as detailed in \cite{Wakeford2020}, which we describe briefly here. We started with the calibrated \texttt{flt} FITS files provided by MAST, which have been processed by the \texttt{calwfc3} pipeline to correct for dark current, read noise, and non-uniform pixel response. In contrast to WFC3/IR observations, for which the first orbit is strongly affected by systematics and must be discarded, WFC3/UVIS observations of the first orbit can be retained. However, previous works have found that when using the WFC3/UVIS, the first frame of each orbit may present unusual high background values and value distributions, with an unknown cause suspected to be earthshine and/or systematics \citep{Wakeford2020}. Following their procedure, we discard each orbit's first frame. We then rejected cosmic rays and hot pixels from each frame in time and space. Cosmic rays were rejected from each pixel through a double iteration process over each pixel's time series. On each iteration, $4.5\sigma$ outliers in each time series were replaced with the median value of that pixel over the past 3 and next 3 frames, ensuring that in-transit outliers were replaced with in-transit values, not with out-of-transit signal. The threshold $4.5\sigma$ was determined by testing threshold values and selecting the lowest value that still produced a random spatial distribution of flagged outliers, since cosmic rays should hit the detector in a random way. About 8\% of pixels were affected by this process. Hot pixels were rejected spatially through one iteration of Laplacian edge detection following \cite{vanDokkum2001}, with rejected pixels being replaced by the median of the pixels one space above, below, left, and right of it. We lastly removed the background sky signal by computing the mode count of each full frame and subtracting that value from the entire frame. As an example, we show in Figure~\ref{fig:frame} one of our time-series frames after the above cleaning procedures have been carried out.

Trace fitting was performed using the \GRISMCONF software developed by \cite{PirzkalRyan2017}, with the latest UVIS/G280 configurations provided by \cite{Pirzkal2020}. The direct image of WASP-127 obtained through the F300X filter was used to predict the position of the dispersed traces in UVIS/G280 images. From the position of the target in the F300X image, \GRISMCONF fit a dispersion polynomial to the $\pm$1st-, $\pm$2nd-, $\pm$3rd-, and $\pm$4th-order traces in the UVIS/G280 images. The extracted stellar spectra contain H$\alpha$ and H$\beta$ absorption lines, from which we could confirm the accuracy of the dispersion polynomial to $\pm$1~nm, just under 1 detector resolution element. \GRISMCONF was used to assign a wavelength solution to the dispersion polynomial, which spans from 200~nm to beyond 800~nm; however, we truncated our analysis to the UVIS span from 202.5~nm to 792.5~nm. The lower bound was set by SNR constraints, while the upper bound was determined by the onset of fringing effects that make reliable extraction impossible. 

UVIS/G280 produces higher-order traces that contaminate the lower-order traces at long wavelengths. In particular, the $\pm$2nd-order traces overlap with the $\pm$1st-order trace, with the sharply-curving UV edge of the 2nd-order trace overlapping the 1st-order trace around 400~nm. Self-contamination is thus present in all wavelengths redder than 400~nm. The severity of self-contamination is target-dependent, being more significant for UV-bright host stars. In the case of WASP-127b, error due to photometric uncertainty was greater than error due to self-contamination, such that we treated self-contamination as negligible. However, we note that for stars with more significant UV emission, self-contamination may require a more sophisticated treatment; we leave further discussion of how to treat this problem for other targets in the HUSTLE program.

We extracted our spectrum from an aperture that we defined as centered on the \GRISMCONF polynomial fit and including all pixels within one half-width $x$ above and below the fit. The preferred half-width should maximize signal while minimizing scatter accumulated from the background. This half-width was determined as follows:
\begin{enumerate}
    \item Define the extraction aperture to be centered on the \GRISMCONF dispersion polynomial fit and containing all pixels that are within one half-width $x$ above and below the fit.
    \item Perform simple unweighted extractions of the $\pm$1st-order traces with apertures of half-width $x$ ranging from 5 to 20 pixels, and sum to produce broad-band light curves.
    \item Fit a simple systematic model (linear trend + exponential ramp) to the out-of-transit flux and measure the scatter.
    \item Select the half-width $x$ that minimizes the out-of-transit scatter.
\end{enumerate}
For this case, $x=10$ was determined to be ideal.

\begin{figure}[h!]
    \centering
    \includegraphics[width=0.48\textwidth]{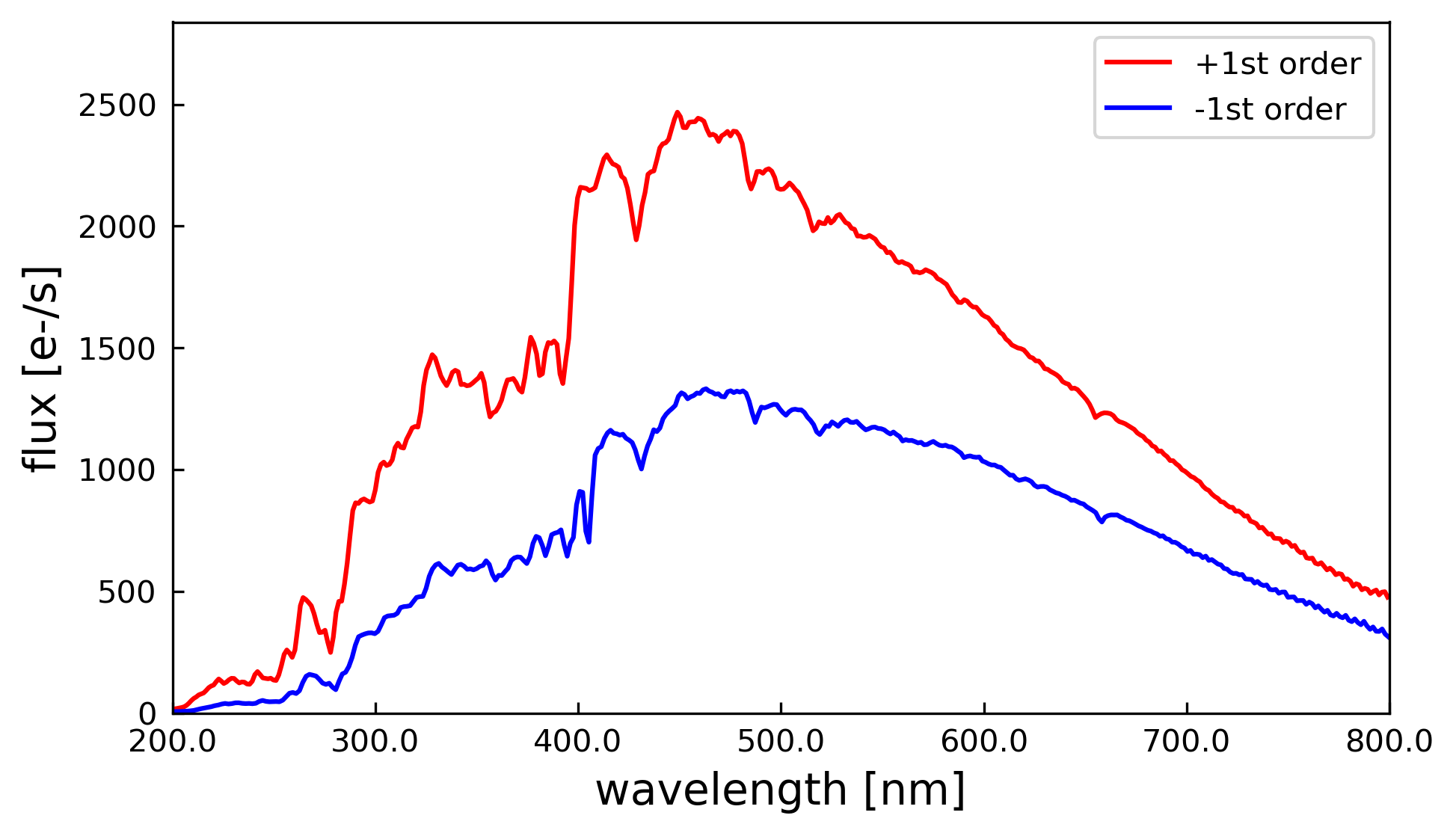}
    \caption{$+1$st and $-1$st-order spectra obtained from our observations of WASP-127b through HST's WFC3/UVIS G280 grism. For the $-1$st order, there is insufficient flux short of 250~nm to perform a reliable analysis. Additionally, an unknown systematic between $-$1st-order wavelengths 397.5~nm and 407.5~nm created steep, spurious and untreatable jumps in flux level, which resulted in this region being excluded from further analysis.}
    \label{fig:1Dspec}
\end{figure}

Pointing instabilities in HST during observations lead to shifts of order tenths of pixels of the trace position on the detector over the course of a visit; however, the wavelength solution is fixed by the F300X direct image, leading to offsets in the wavelengths assigned to each frame's 1-dimensional (1D) spectrum. To account for this offset, all 1D spectra were cross-correlated with the 1D spectrum extracted from frame 2 of orbit 1. Through cross-correlation, we measured spectral displacements and re-aligned all spectra to the frame 2 spectrum to ensure a consistent wavelength solution. A final round of cosmic ray cleaning was then performed by comparing all spectra to the median spectrum and replacing $3.5\sigma$ outliers with the median value in time. Under 0.2\% of 1D spectral points were deemed outliers at this stage with 72 total residual hits across both orders, as to be expected after undergoing the various cleaning treatments detailed above. Two extracted 1D spectra, one from the $+1$st order and one from the $-1$st order and both processed as described above, are presented in Figure~\ref{fig:1Dspec}.

We chose to extract our spectra from the $+1$st- and $-1$st-order traces as these orders have the highest SNR and widest wavelength coverage, while being the least affected by self-contamination from other spectral orders. We summed each order of 1D spectra across its entire spectral range to produce two broad-band light curves, one per order. Spectroscopic light curves were produced from the 1D spectra by summing each spectrum across discrete wavelength bins with bin widths of 10~nm. From the brighter $+1$st order, 59 spectroscopic light curves spanning 202.5~nm to 792.5~nm were produced. The $-1$st order is considerably dimmer below 250~nm and had additional systematics between 397.5~nm and 407.5~nm, so only 52 spectroscopic light curves spanning 250~nm to 397.5~nm and 407.5~nm to 792.5~nm were produced from this order. In total, 111 spectroscopic light curves and two broad-band light curves were produced for analysis, the details of which are provided in Section~\ref{sec:analysis}.

\subsubsection{\lluvia} \label{sec:P2}
\lluvia is being developed independently of \Hazelnut and is detailed thoroughly in \citet[][in prep.]{Carlos}; we provide a brief summary of its methods here along with its results. Our \lluvia reduction likewise began with the \texttt{flt} FITS provided by MAST and calibrated by \texttt{calwfc3}, but retained all frames in each orbit including the first frames, in contrast to \Hazelnut. The background flux of each exposure was computed using four 150~$\times$~400 pixel regions located in the four corners of the corresponding image. The central value of the histogram of counts across all four regions was then used as the image background. An initial cosmic ray correction was performed on each image by iteratively identifying pixels that deviated by more than 5$\sigma$ in time along each HST orbit, and replacing their flux values by the median pixel value of the corresponding orbit.

We calculated the position of both $\pm$1 order spectral traces by fitting a Gaussian curve to each detector column (i.e., cross-dispersion direction), using the calibration trace obtained from the \GRISMCONF package as an initial guess. We extracted the stellar spectra using the optimal extraction routine described in \cite{Horne1986}, with a $\pm10$-pixel aperture around the center of the trace. To account for the curvature present in the traces of WFC3/UVIS G280, we built the spatial profile by fitting low order polynomials along the trace direction, similar to the methodology proposed in \cite{Marsch1989}. We used this fitting process to perform a second cosmic ray removal; pixels which deviated from the spatial profile model by more than $7\sigma$ were replaced by the corresponding value of the polynomial fit. We integrated the flux across the entire wavelength range to produce the $\pm$1st-order broad-band light curves. We also divided the extracted stellar spectra into 10~nm bins, to produce a set of 57 spectral light curves for both $\pm$1 orders. The analysis of the broad-band and spectroscopic light curves produced here is also detailed in Section~\ref{sec:analysis}.

\section{Light Curve Fitting} \label{sec:analysis}

\subsection{\Hazelnut}
\subsubsection{Systematics Removal}
Our \Hazelnut broad-band and spectroscopic light curves were largely cleaned of outliers but were still affected by HST systematics, some of which were wavelength-dependent. Thus, each light curve had to be individually treated for HST systematics. We did so through the \xISM package \citep{Exoticism}, the procedure behind which is detailed in \cite{Wakeford2016}. \xISM fit a weighted combination of 51 systematics models to remove all known types of systematic trends found in our data, including HST thermal breathing, visit-long slopes, and ramp effects. We used \xISM to detrend all of our broad-band and spectroscopic light curves before proceeding with further analysis.

\subsubsection{Broad-band Light Curve Analysis}
\begin{figure*}[ht!]
     \centering
     \includegraphics[width=\columnwidth]{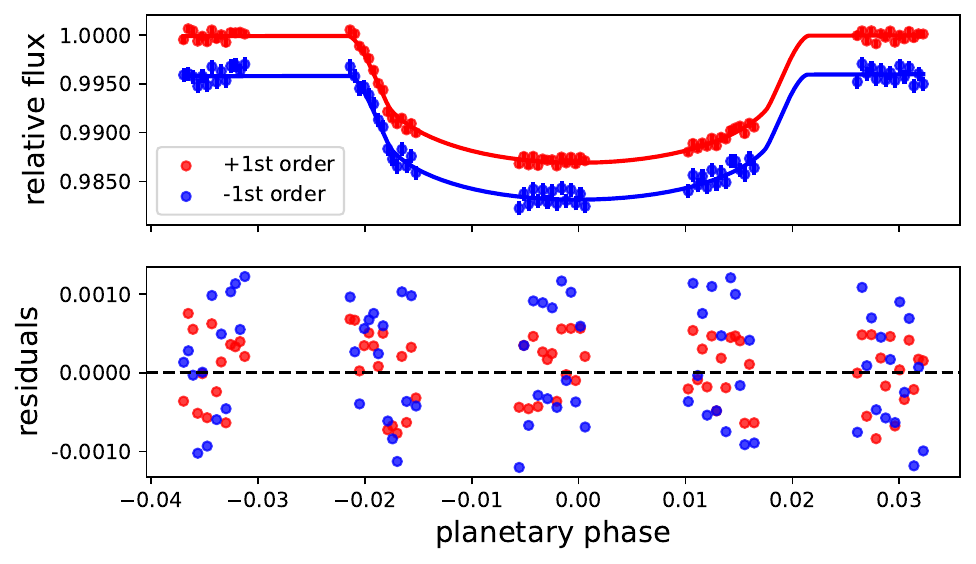}
     \includegraphics[width=\columnwidth]{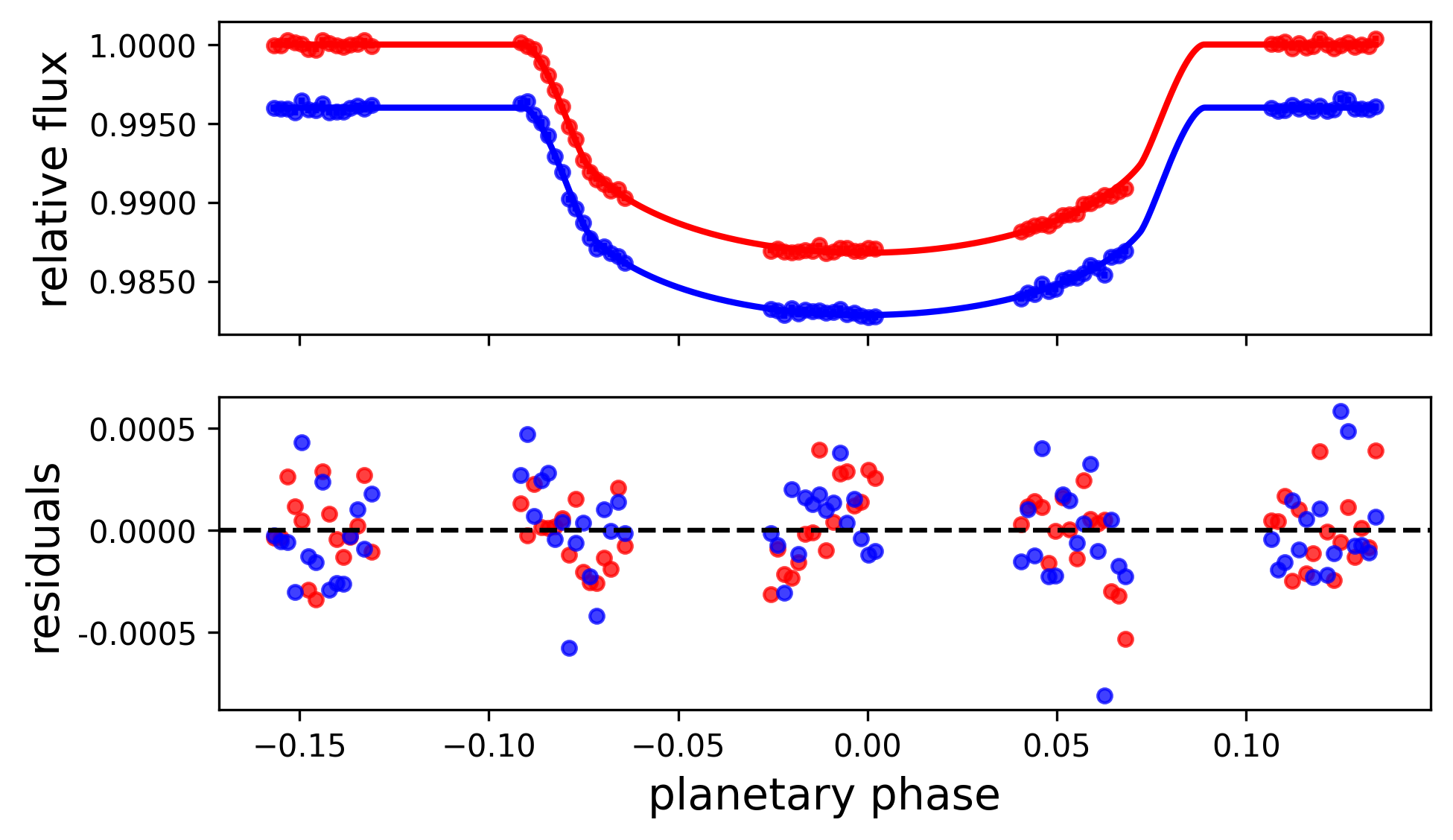}
    \caption{Top: Normalized and detrended broad-band light curves with fitted models overplotted. Bottom: Residuals after fitting. $+1$st order light curves and residuals are in red, while blue denotes the $-1$st order. Left: The broad-band light curves and residuals obtained by \Hazelnut. For clarity, a 400~ppm offset is applied to the $-1$st order. A ``doubling'' effect is seen in the light curves, most prominently in the $-1$st order. This effect arises from the method by which background values in each frame were selected; this doubling does not notably affect the results. Right: The broad-band light curves obtained by \lluvia. The broad-band precision achieved by \lluvia is greater; nevertheless, both pipelines obtain comparable median spectral precision and spectra largely consistent to 1$\sigma$.}
    \label{fig:WLC}
\end{figure*}

We first analyzed our detrended broad-band light curves spanning the entire spectral range of 202.5~nm to 792.5~nm. For each broad-band light curve, we fit a \cite{MandelAgol2002} transit model implemented through the \texttt{batman} Python package \citep{Batman}, with a four-parameter nonlinear stellar limb darkening law. Fixed limb darkening coefficients were generated by \xLD \citep{Exotic2022} using the ATLAS9 grid \citep{Kurucz1993}. Our choice of 1D stellar model grid was motivated by consistency of comparison with \citet{Spake2021}, but we note that our secondary reduction with \lluvia uses a 3D grid instead. The stellar parameters we supplied to \xLD are sourced from \citet{Lam2017} and reported in Table~\ref{tab:params}. Our fitting procedure is a two-step process that uses both linear and non-linear fitting techniques. First, we used a \texttt{scipy} linear least-squares solver to fit a first-pass transit model, which includes a visit-long linear-in-time systematic trend. This model was used to revise our estimates of the photometric uncertainty as well as fit for the mid-transit epoch $T_0$, the ratio of the semimajor axis to the host star radius $a/R_*$, the impact parameter $b$, and the slope and y-intercept of the systematic. We then inputted the uncertainty estimates, systematic model, and the fitted values of $T_0$, $a/R_*$, and $b$ to a routine that used \texttt{EMCEE} \citep{Emcee} to refit the transit model through a Markov chain Monte Carlo (MCMC) algorithm with uninformed priors. We initialized our MCMC routine with planetary parameters from \citet{Lam2017} and ran 32 chains with 5000 steps each, discarding the first 640 steps (20\% of each chain) as burn-in. We present a summary of the system parameters in Table~\ref{tab:params}, and in the left-side panels of Figure~\ref{fig:WLC} we present the marginalized broad-band light curves with fitted models and normalized residuals for each curve.

Our broad-band light curves yielded a transit depth of $0.997\pm0.009\%$ in the $+1$st order and $0.988\pm0.016\%$ in the $-1$st order, for an overall depth precision of 91~ppm. Our $+$1st-order result is notably shallower than the G430L depth observed by \citet{Spake2021} ($1.034^{+0.006}_{-0.005}\%$), within 2$\sigma$ of their G750L depth ($1.013^{+0.009}_{-0.006}\%$) and agrees very well with their G141 depth ($0.996\pm0.011\%$). We ascribe these differences primarily to changes in instrument systematics between visits, and apply offsets to the \citet{Spake2021} spectrum to align it with our own during the interpretation in Section~\ref{sec:retrieval}. The $-1$st-order depth is notably more uncertain, as expected, since the negative orders are roughly half as bright as the positive orders. We fit system parameters separately for each order and combined our results in a weighted average, yielding orbit semimajor axis and impact parameter $a/R_*=8.116\pm0.106$ and $b=0.127\pm0.062$, consistent to $1\sigma$ with the values presented in \citet{Lam2017}. Despite being fit separately, our $+1$st- and $-1$st-order fit parameters were also consistent to well within $1\sigma$ with each other. Our mid-transit epoch was determined to be $T_0=2459989.546\pm0.001$ in $\mathrm{BJD}_{\mathrm{TDB}}$. In our subsequent analysis of the spectroscopic transit curves, we fixed $a/R_*$, $b$, and $T_0$ to the values obtained here.

\begin{table}[]
    \centering
    \begin{tabular}{l|l|l}
    \hline
    \hline
    Parameter & Value & Source \\
    \hline
    Stellar parameters & & \\
    \hline
    $Z$ [Fe/H] & $-0.18\pm0.06$ & \citet{Lam2017} \\
    $T_{\mathrm{eff}}$ [K] & $5620\pm85$ & \citet{Lam2017} \\
    $\log(g)$ [cgs] & $4.18\pm0.01$ & \citet{Lam2017} \\
    \hline
    Exoplanet \\ Parameters & \\
    \hline
    $R_p/R_*$ & $0.1000\pm0.0004$ & This work \\
    $a/R_*$ & $8.116\pm0.106$ & This work \\
    $b$ & $0.127\pm0.062$ & This work \\
    $e$ & $0.0$ & This work \\
    $P$ [days] & $4.178062$ & \citet{Spake2021} \\
    $t_0$ [$\mathrm{BJD}_{\mathrm{TDB}}$] & $2459989.546\pm0.001$ & This work \\
    \hline
    \hline
    \end{tabular}
    \caption{System parameters used for fitting broad-band and spectroscopic transit light curves. Stellar parameters were fixed from \cite{Lam2017}. Exoplanet parameters were fitted by this analysis and derived from a weighted average of the MCMC fitting results from each order, except for the period $P$, which was fixed by \cite{Spake2021} and the eccentricity $e$, which we took to be $0.0$.}
    \label{tab:params}
\end{table}

\subsubsection{Spectroscopic Light Curves Analysis}
We followed a very similar analysis procedure for the spectroscopic light curves, first using a linear least-squares fit to update the photometric uncertainty estimates and then using MCMC to derive a final fit with uncertainty estimates on the fitted parameters. We fit a transit model with a linear-in-time systematic and fixed limb darkening coefficients generated for each wavelength bin by \xLD with a four-parameter nonlinear limb darkening law. As before, ATLAS9 grid coefficients were supplied by \xLD and stellar parameters were sourced from \citet{Lam2017}. Our spectroscopic light curves were fitted with the planet radius-to-stellar radius ratio $R_p/R_*$ as the sole free physical parameter, as well as fitting a slope and y-intercept for a wavelength-dependent systematic trend.

Following the procedure of \cite{Wakeford2020}, we extracted transit depths $(R_p/R_*)^2$ at each wavelength from each order and averaged the results from the two orders together, excluding $-1$st-order wavelengths bluer than 250~nm because the $-1$st order mid-UV flux is too low for reliable analysis. We additionally excluded the $-1$st-order wavelengths 397.5~nm and 407.5~nm which, as stated in Section~\ref{sec:data}, suffered an unknown systematic that could not be removed by \xISM. We present our final \Hazelnut spectrum and limb darkening coefficients in Appendix A, Table~\ref{tab:spectrum} as well as in Figure~\ref{fig:spectrum} alongside our \lluvia reduction and spectra obtained by other observations. Our median spectroscopic precision in our 10~nm bins is $\sim240$~ppm.

To benchmark the performance of the WFC3/UVIS G280 grism in comparison to WFC3/IR and STIS gratings, we also re-binned our \Hazelnut 1D spectra to match the wavelength binning scheme used in \citet{Spake2021}. To produce a near one-to-one comparison, we fit our re-binned light curves following the same limb darkening model and system parameters as used by \citet{Spake2021}. Our reduction, extraction, and fitting techniques are otherwise unchanged from the analysis detailed above; we present the results of this comparative analysis in Figure~\ref{fig:spectrum} and discuss its implications in Section~\ref{sec:disc}.

\subsection{\lluvia}
\subsubsection{Systematics Removal}
As with the light curves produced by \Hazelnut, the light curves produced by \lluvia were also affected by HST systematics. We modeled each light curve $f(t)$ with a function of the form $ f(t) = F_{0} \times T(t, \theta) \times S(x) $, where $F_{0}$ is the baseline flux, $T(t, \theta)$ is the analytic transit model implemented through the \texttt{batman} package \citep{Batman} (which is a function of the system parameters $\theta$), and $S(x)$ is an instrument systematics model. Following the jitter decorrelation method described in \cite{Sing2019}, we defined $S(x)$ as a linear combination of several detrending variables. These include optical state vectors estimated from the observations (e.g., HST orbital phase, displacements along the X and Y detector directions, etc.), and jitter vectors extracted from the HST Pointing Control System (e.g., telescope right ascension, declination, etc.).

\subsubsection{Broad-band Light Curve Analysis}
We also started by analyzing the broad-band light curves, fitting for the mid-transit epoch $T_{0}$, planet-to-star radius ratio $R_p/R_*$ and baseline flux $F_0$. However, we did not fit for system parameters $a/R_*$ and $b$; instead, we fixed these values to those obtained by the \Hazelnut fit for consistency of comparison. Furthermore, we adopted the four-parameter non-linear limb darkening law as implemented in \xLD \citep{Exotic2022}, using the 3D stellar models from \cite{Magic2015}, in contrast with our \Hazelnut fit which uses a 1D stellar model grid. To avoid over-fitting, we tested all the possible combinations of the detrending variables included in the instrument systematics model $S(x)$, and selected the subset of detrending variables that maximized the second-order Akaike Information Criterion (AIC$_c$). For all light curves, we first performed a linear least squares fit of the $f(t)$ model to reject $3.5\sigma$ outliers and rescale the error bars to achieve a reduced-$\chi^2$ of unity. We then re-fitted the $f(t)$ model through MCMC with \texttt{emcee} \citep{Emcee}. We present the detrended and fitted broad-band light curves in the right-side panels of Figure~\ref{fig:WLC}.

\subsubsection{Spectroscopic Light Curve Analysis}
We fitted each of the spectroscopic light curves following a similar procedure as in the broad-band light curve analysis, but fixing the mid-transit epoch $T_{0}$ to the value extracted from the broad-band light curves, and fitting only for the planet-to-star radius ratio $R_p/R_*$ and the baseline flux $F_0$. Furthermore, we constructed our systematics model $S(x)$ using the best-fitting subset of detrending variables derived from the broad-band analysis. We extracted two independent transmission spectra for orders $+1$ and $-1$, which we subsequently combined with a weighted mean to produce the final transmission spectrum. As in \Hazelnut, due to the poor precision in the bluer end of the $-1$st-order, we excluded the $-1$st-order light curves below 250 nm. The final weighted spectrum is presented in panel 2 of Figure~\ref{fig:spectrum}.

\subsection{HST WFC3/UVIS G280 Spectrum}
\begin{figure}[h!]
    \centering
    \includegraphics[width=\columnwidth]{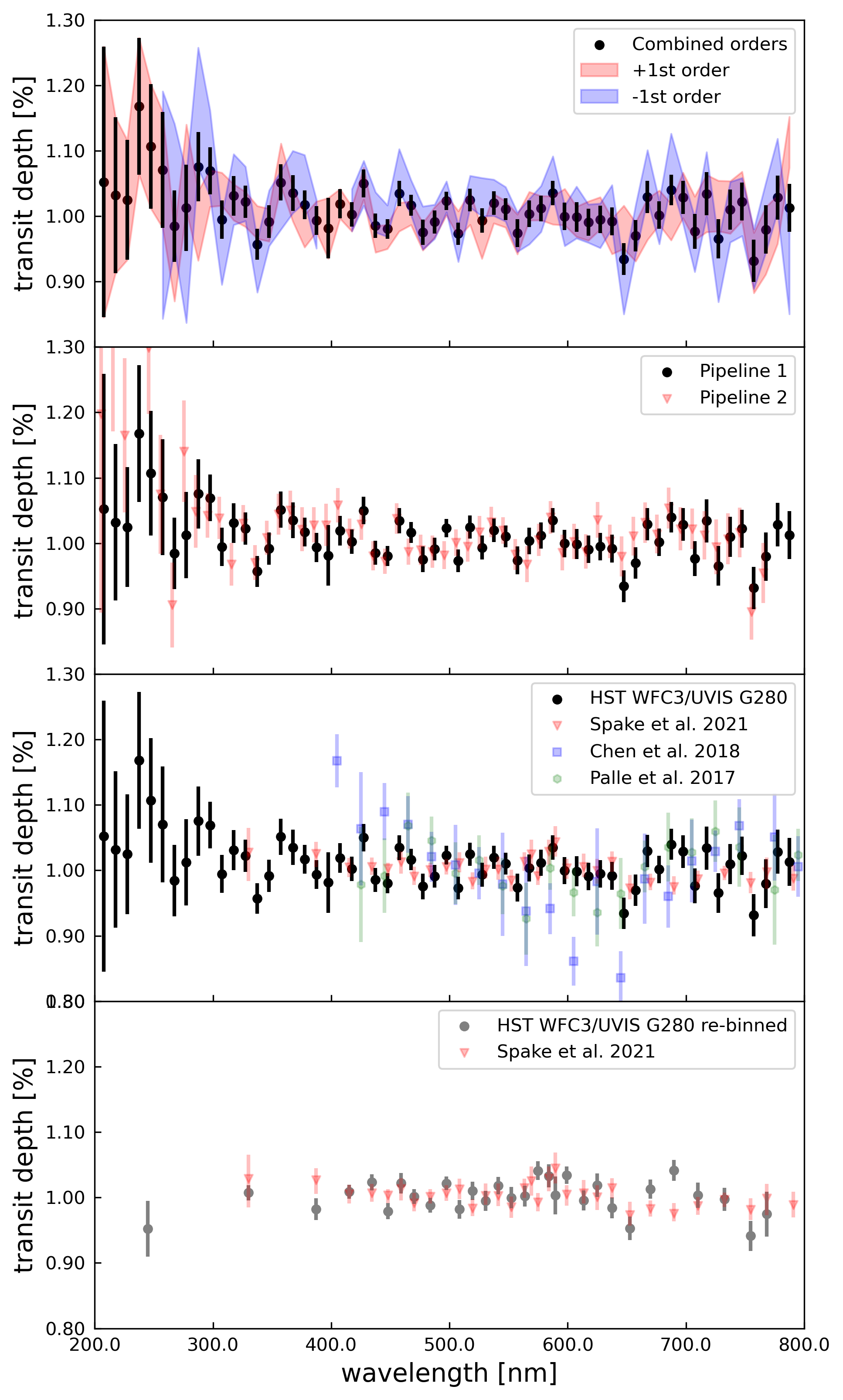}
    \caption{Spectra of WASP-127b obtained from this and other analyses. Panels are numbered 1-4 from the top. Panel 1: The \Hazelnut reduction of WASP-127b's HST WFC3/UVIS G280 observations shows both the $+1$st and $-1$st orders both separately and combined. Transit depths below 250~nm and in between 397.5 and 402.5~nm from the $-1$st order are excluded for data quality reasons. Panel 2: Both reductions of our UVIS observations with no offsets applied, showing agreement to within 1$\sigma$ for most points, with a steeper mid-UV spectrum obtained by the \lluvia reduction that nonetheless remains within $1\sigma$ uncertainties. Panel 3: The \Hazelnut reduction compared to the spectra obtained by \cite{Palle2017}, \cite{Chen2018}, and \cite{Spake2021}, with offsets of +236, -1050, and -222 ppm respectively applied to account for discrepancies arising from different detectors and observing epochs. Panel 4: The \Hazelnut spectrum re-binned to provide a near one-to-one comparison to \citet{Spake2021}, with a -212 ppm offset applied to the latter spectrum.}
    \label{fig:spectrum}
\end{figure}

We present the UVIS spectra obtained by \Hazelnut and \lluvia in Figure~\ref{fig:spectrum}. We showcase the $+1$st and $-1$st order separately for \Hazelnut (Figure~\ref{fig:spectrum}, panel 1), and compare our \Hazelnut spectrum to our \lluvia spectrum (Figure~\ref{fig:spectrum}, panel 2) as well as to previous measurements of WASP-127b's spectrum (Figure~\ref{fig:spectrum}, panel 3). We find that while both UVIS/G280 pipelines were developed and operated mostly independently and used different methods of cleaning and fitting the data, the majority of transit depths in both spectra agree to within $1\sigma$. We also find our \Hazelnut spectrum with uniform 10~nm bins and independently-determined system parameters to be in good agreement with the STIS spectra obtained by \citet{Spake2021}, recovering a similar shallow slope as well as a small absorption peak near 670~nm corresponding to a well-known sodium absorption feature. The larger uncertainties in the spectra of \citet{Palle2017} and \citet{Chen2018} means that both are consistent with our UVIS/G280 spectrum to within 1--2$\sigma$; however, the apparent strong slope in \citet{Chen2018} is not seen in either the STIS spectrum nor the UVIS/G280 spectrum. We discuss this further in Section~\ref{sec:disc}. Our re-binned \Hazelnut spectrum (Figure~\ref{fig:spectrum}, panel 4), which uses the same wavelength binning scheme, system parameters, and limb darkening as \citet{Spake2021}, also shows reasonable agreement with \citet{Spake2021}, with only a few notable discrepancies in the redder wavelengths and notably higher precision, which will also be discussed further in Section~\ref{sec:disc}.

\section{Forward Models and Atmospheric Retrievals} \label{sec:retrieval}

\subsection{Forward Modeling with \texttt{PICASO} and \texttt{Virga}}
Forward modeling was carried out on the primary reduction spectrum from \Hazelnut shown in Figure~\ref{fig:spectrum} and tabulated in Table~\ref{tab:spectrum}. We utilized \texttt{PICASO} for both radiative transfer \citep{PICASO2019} and self-consistent climate modeling \citep{PICASO2023}. \texttt{PICASO} is a radiative-convective thermochemical equilibrium (RCTE) model that allows the user to model the pressure-temperature (P-T) profile and 1D transmission or emission spectrum of an exoplanet or brown dwarf's atmosphere based on the object's temperature structure, gravity, and host star characteristics. 

\texttt{PICASO} self-consistently models equilibrium chemistry and clouds, the latter of which is made available by coupling \texttt{PICASO} to \texttt{Virga} \citep{virga_initialrelease, virga_details}, a program for modeling the distribution and composition of clouds in an exoplanet's atmosphere based on the methods of \texttt{eddysed} by \cite{virga_methods}. \texttt{Virga} provides a physically-motivated estimate of the cloud contribution by including the wavelength-dependent opacity according to the particle size distribution. The aerosol size distribution is determined by the balance of cloud sedimentation efficiency ($f_{\rm{sed}}$) against upward vertical mixing as parameterized by the eddy diffusivity ($K_{\rm{zz}}$) for condensing species.

By fitting a distribution of \texttt{PICASO} forward models to our WASP-127b transmission spectrum, inferences about the composition and structure of its atmosphere can be made. Our \texttt{PICASO} grid consists of atmospheric compositions set by the RCTE calculations of \citet{Marley2021} using correlated-$k$s of \citet{LupuCK_picaso}. These correlated-$k$ opacities include \ce{C2H2}, \ce{C2H4}, \ce{C2H6}, \ce{CH4}, CO, \ce{CO2}, CrH, Fe, FeH, \ce{H2}, \ce{H3+}, \ce{H2O}, \ce{H2S}, HCN, LiCl, LiF, LiH, MgH, \ce{N2}, \ce{NH3}, OCS, \ce{PH3}, SiO, TiO, VO, Li, Na, K, Rb, and Cs. Our grid includes metallicities of 1$\times$, 5$\times$, 10$\times$, 20$\times$, 30$\times$, 50$\times$, and 100$\times$ solar and C/O ratios of 0.115, 0.229, 0.458, and 0.687, which correspond to 0.25$\times$, 0.5$\times$, 1$\times$, and 1.5$\times$ solar C/O following \citet{Lodders2009}. 

For the P-T profile, we initiate the profile with an initial estimate of the temperature structure from the parameterization of \citet{Guillot2010}, using an intrinsic temperature of 200 K \citep[e.g.,][]{Thorngren2019} and a heat redistribution factor of 0.5. Previous works studying planets of a similar equilibrium temperature to WASP-127 b \citep[WASP-39 b, WASP-17b;][]{Ahrer2023,Alderson2023,Feinstein2023,Rustamkulov2023,Grant2023} have found negligible effects on cloud constraints by changing the intrinsic temperature within reasonable values \citep{Thorngren2019} for a planet like WASP-127b. Our observations are also only sensitive to around 0.1 bar, and changing the intrinsic temperature would largely only change the pressure-temperature profile deeper than a bar. After initializing our profile, we then iterate to locate the extent and placement of convective zones until convergence is reached. We inspect each computed cloud-free P-T profile for convergence before proceeding to post-processing clouds.

\begin{figure}[h!]
    \centering
    \includegraphics[width=\columnwidth]{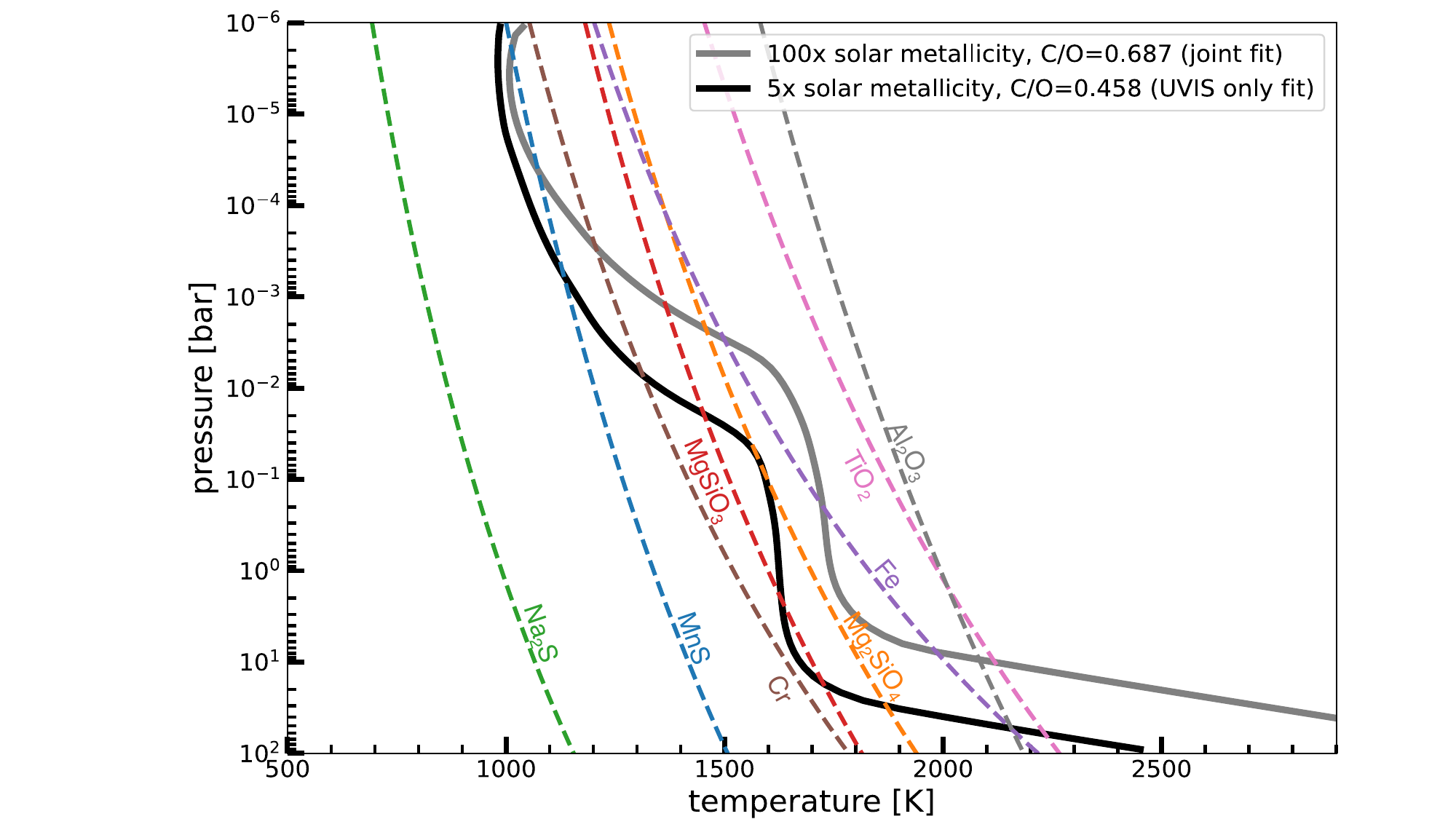}
    \caption{Pressure-temperature profiles computed by \texttt{PICASO 3.0} for WASP-127b, fitted with WFC3/UVIS G280 alone (black) and with the STIS G430L+G750L and WFC3/IR G141 spectra obtained by \cite{Spake2021} (grey).
     Condensation curves at solar metallicity for typical cloud-forming species in hot Jupiter atmospheres (e.g. MgSiO$_3$, MnS) are also plotted; where the P-T profile crosses left of these curves, clouds of that condensate species may form.}
    \label{fig:PT}
\end{figure}

Using \texttt{Virga}, we added aerosol opacity accounting for clouds of MnS, \ce{MgSiO3}, and \ce{Mg2SiO4} with $f_{\rm{sed}}$ of 0.3, 0.6, 1, 3, 6 and $K_{\rm{zz}}$ of 10$^7$, 10$^9$, 10$^{11}$, 10$^{13}$ cm$^2$~s$^{-1}$. We account for atmospheric metallicity in our condensation of these cloud species.  The range of metallicities we consider in our forward model grid produces varying pressure-temperature profiles and molecular abundances that also produce a wide range of potential cloud masses and locations to compare against our observations. In Figure~\ref{fig:PT} we present best-fits of our \texttt{PICASO} P-T profiles which are themselves cloud-free. We also show the condensation curves of cloud species shown under solar metallicity for illustration, demonstrating that the clouds we include in our \texttt{Virga} runs are possible to form in a variety of expected scenarios for WASP-127b. Though they are relevant given our temperature regime as shown in Figure~\ref{fig:PT}, we do not include Fe or Cr clouds as it is expected to be difficult to nucleate and form such clouds compared to the silicate species \citep{Gao2020}. Furthermore, any Fe or Cr clouds that did form would do so at depth, below observable levels. While MnS is also expected to be sluggish to nucleate \citep{Gao2020}, the fact that its condensation curve crosses our temperature-pressure profiles at $\sim$millibar levels led us to include it in our model runs. \ce{SiO2}'s condensation curve lies between \ce{MgSiO3} and \ce{Mg2SiO4}'s \citep{Grant2023}, and so we do not model it separately as its contribution to our model spectra within Hubble's sensitivity is captured by the other two silicates.

To fit our \texttt{PICASO} forward models to the WFC3/UVIS G280 data, we generate synthetic transmission spectra from our converged climate profiles using \texttt{PICASO}'s radiative transfer module. For the radiative transfer, we use an opacity database containing the same species as above from 0.2 to 3 \textmu m resampled to R=10,000 from original line-by-line calculations at R$\sim$1$\times$10$^6$ performed by \citet{Freedman2008} and \citet{Ehsan2021}. We bin the resulting models to the spacing of the data and compute a reduced-$\chi$$^2$ to evaluate the goodness of fit. We consider 3 fits: 1) using the WFC3/UVIS G280 data of this work alone (59 degrees of freedom), 2) the \citet{Spake2021} HST data from STIS G430L+G750L and WFC3 G141 (61 degrees of freedom), 3) and a joint fit to the UVIS/G280 data combined with the HST/STIS G430L+G750L and HST/WFC3 G141 data (120 degrees of freedom). We allow for an offset of the \citet{Spake2021} combined HST reduction compared to our UVIS data. For our fit of the HST/STIS G430L+G750L and HST/WFC3 G141 data (61 degrees of freedom), we confirm that we recover a similar solution to that of \citet{Spake2021}.

\begin{figure*}[ht!]
    \centering
    \includegraphics[width=0.8\textwidth]{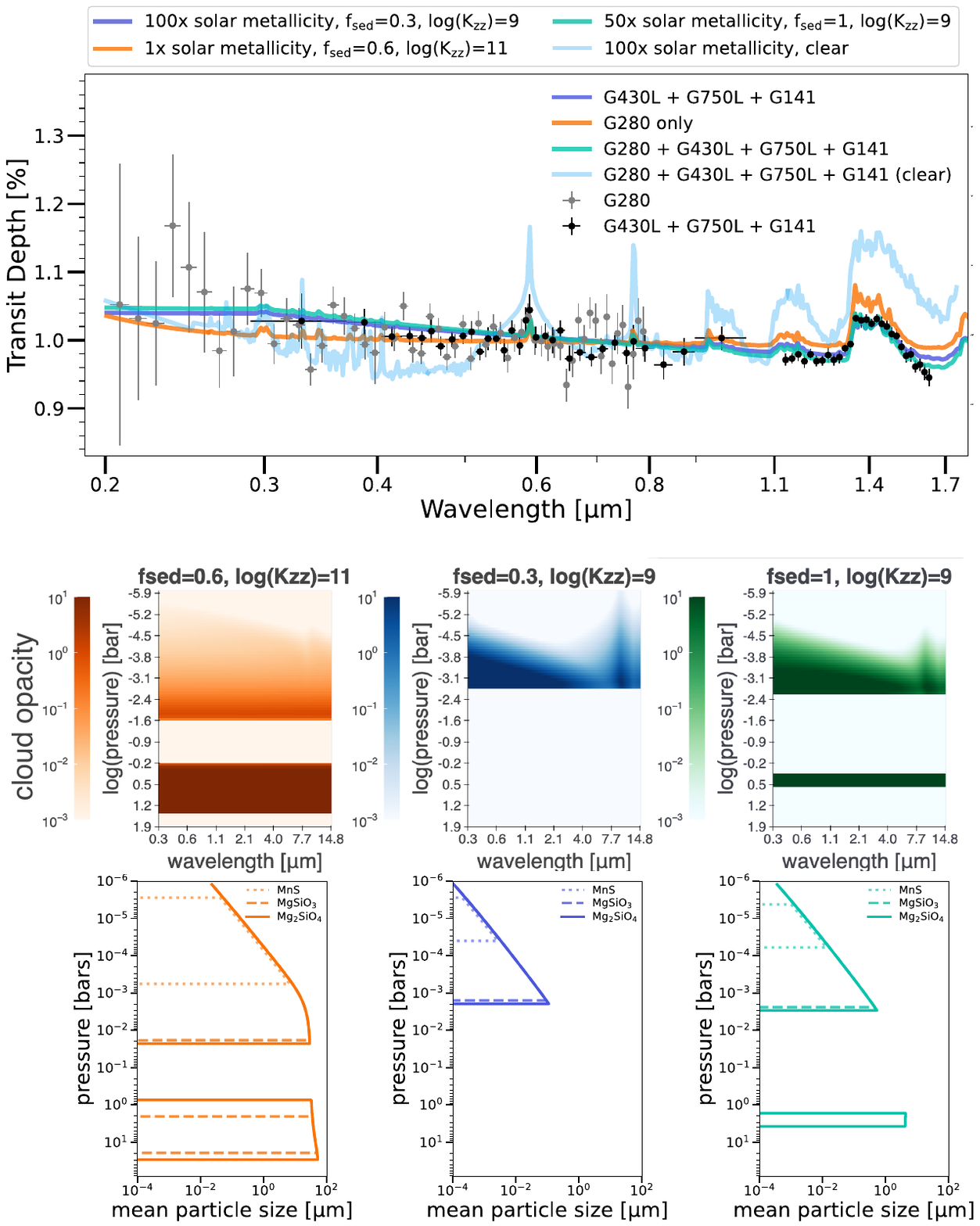}
    \caption{Top: \texttt{PICASO} best-fit forward models of the UVIS/G280 spectrum of WASP-127b (grey points), fitted both separately and jointly with the STIS G430L+G750L and WFC3/IR G141 spectra obtained by \cite{Spake2021} (black points). The best fit of a clear jointly fitted model is also shown in light blue, demonstrating that clouds are required to fit the spectrum well. Middle: Cloud opacities for the single best-fitting \texttt{Virga} runs for the UVIS/G280 spectrum (orange), STIS/G141 spectrum (blue), and joint fit (green). Bottom: Mean particle sizes for each corresponding \texttt{Virga} run in the row above.}
    \label{fig:models}
\end{figure*}

We show a selection of our forward model fits in Figure~\ref{fig:models}. Our clear model atmospheric spectra are all poor fits to the data, with a best fit reduced-$\chi$$^2$=2.3 for the UVIS/G280 only fit (59 DOF), reduced-$\chi$$^2$=24.7 for the STIS G430L+G750L+WFC3 G141 fit (61 DOF), and reduced-$\chi$$^2$=24.5 (120 DOF) for the joint UVIS/STIS/G141 fit. Our cloudy fits comparatively provide reduced-$\chi$$^2$=1.2, 0.89, and 0.81, for UVIS, STIS/G141, and UVIS/STIS/G141 respectively. Our forward model grid tops out at 100$\times$ solar metallicity, which is expected to be a reasonable upper limit given the known constraints from JWST for a variety of giant planet atmospheres \citep{KemptonKnutson2024}. Given this 100$\times$ solar metallicity limit, we are unable to generate clear atmospheres with small enough amplitudes at WFC3/G141 wavelengths to obtain good fits to the data, driving these large reduced-$\chi$$^2$s for the clear runs. Conversely, adding cloud opacity easily suppresses the amplitude of the water feature present in WFC3/G141, resulting in the much better reduced-$\chi$$^2$s for the cloudy models.

The clear atmosphere preferred models are all uniformly 100$\times$ solar metallicity, while cloudy models reduce the best-fit preferred metallicity down to 1$\times$ solar for the UVIS/G280 data alone, 100$\times$ solar for the STIS/G141 data alone, and 50$\times$ solar for the joint fit. However, this is not strongly constrained by our \texttt{PICASO} forward models, and the degeneracy between metallicity and cloud parameters means that any number of combinations provide a good fit to the data. For example, we find that our goodness-of-fits for the joint dataset (G280 + STIS + G141; 120 DOF) range from 5--100$\times$ solar metallicity across the full range of cloud parameters with reduced-$\chi$$^2$ from 0.8--2.0, which is not a statistically significant interval. Nonetheless, the significant difference in fits between our clear and cloudy runs suggest that stronger scattering imparted by aerosol particles, combined with the muting of the water feature in the WFC3/G141 bandpass, requires the presence of clouds in our forward models. 

Within our modeling framework, $f_{\rm{sed}}$ parameterizes the efficiency with which particles sediment down through the atmosphere while $K_{\rm{zz}}$ sets the strength of upward vertical mixing. Therefore the balance between $K_{\rm{zz}}$ and $f_{\rm{sed}}$ sets cloud particle sizes, with larger cloud particles in more compact layers produced by large $f_{\rm{sed}}$ and small $K_{\rm{zz}}$, or smaller cloud particles in more extended cloud layers produced by small $f_{\rm{sed}}$ and larger $K_{\rm{zz}}$. Our best-fitting cloudy models range from $K_{\rm{zz}}$=10$^9$--10$^{11}$ cm$^2$ s$^{-1}$ with $f_{\rm{sed}}$ $\leq$ 1. The UVIS/G280 only fits have an $f_{\rm{sed}}$ = 0.6, while the STIS/G141 and joint fits have a wider spread anywhere from $f_{\rm{sed}}$ = 0.3--1.  For the UVIS/G280 cloudy fits, our \texttt{Virga} runs result in micron-sized particles made primarily of silicates, while the STIS/G141 data alone favors cloudy fits with sub-micron sized particles. 

None of the cloud mixing parameters have particularly strong constraints, but when analyzed jointly, the data result in cloudy models with relatively high altitude ($\sim$millibar--microbar), small (sub-micron) cloud particles. All the cloud fits result in a cloud layer dominated by silicate opacity, with only a very optically thin contribution from any MnS cloud. We examined whether MnS clouds alone could match the data, and found that their much smaller cloud mass and correspondingly weaker opacity means that silicate clouds are required to explain the data given our model setup. The cloud scenario for each dataset combination is shown in the lower two rows of Figure~\ref{fig:models}, which highlights both the cloud optical depth (middle row) and the mean effective particle size in the cloud layer by species (bottom row), demonstrating the dominance of silicate cloud compared to manganese sulfide cloud opacity.

\subsection{Retrievals with \texttt{POSEIDON}}


We additionally perform an atmospheric retrieval analysis, in which the parameter space of possible spectra is built by varying different model parameters (e.g., molecular abundances, P-T profiles, and aerosol properties). Retrievals employ Bayesian methods to explore the model parameter space and determine the range of atmospheric properties that provide good fits to the data. For our retrievals, we employ the open source \texttt{POSEIDON} code \citep{MacDonald2017,MacDonald2023}.

\subsubsection{Retrieval Configuration}

We construct a suite of retrievals to assess the contribution of our UVIS/G280 observations to the inferred atmospheric parameters of WASP-127b. Our retrievals consider multiple combinations of datasets, including G280 alone, G280 in concert with archival WASP-127b transmission spectra from \citet{Spake2021} (STIS, WFC3/IR G141, and \textit{Spitzer} photometry), and the archival data alone.

Our retrievals employ a parametric model of a H$_2$-He dominated atmosphere containing multiple trace chemical species. We assume 1D atmospheres (varying only with altitude) with an isothermal P-T profile, parameterized by the isothermal temperature, $T$, and a pressure grid with 100 levels spanning 100--$10^{-9}$\,bar. We parameterize trace chemical species via their $\log_{10}$ volume mixing ratio (VMR), considering atomic and molecular opacities from \ce{Na} \citep{Ryabchikova2015}, \ce{K} \citep{Ryabchikova2015}, \ce{FeH} \citep{wende2010crires}, \ce{H2O} \citep{polyansky2018exomol}, \ce{CH4} \citep{yurchenko2017hybrid}, \ce{CO} \citep{li2015rovibrational}, and \ce{CO2} \citep{tashkun2011cdsd}. \ce{H2} and \ce{He} pressure broadening is applied for all species except FeH (which uses the broadening prescription from \cite{sharp2007atomic}). Our retrievals also consider inhomogeneous aerosols via the 4-parameter prescription from \citet{MacDonald2017}. This parameterization couples optical scattering (defined by two parameters: the power-law exponent of the scattering slope, $\gamma$, and an enhancement over Rayleigh scattering, $\log{a}$) with an opaque grey cloud deck, defined at a cloud top pressure $\log({\mathrm{P_{cloud}}})$, covering a fraction $\phi_{\rm{cloud}}$ of the terminator. For retrievals with multiple datasets, we include a free offset, $\delta_{\mathrm{rel}}$, between the UVIS G280 data and the archival data. STIS, WFC3, and Spitzer were not offset from each other in this analysis as they were jointly analyzed by \citet{Spake2021} who included offsets in their own work. G280 as a separate analysis required the consideration of an offset to the prior jointly analyzed data. We additionally fit for the radius of the planet at the 10\,bar reference level, R$_{p, \, \mathrm{ref}}$. We compute model spectra for each atmosphere on a wavelength grid with R=5,000 from 0.18--5.2\,$\mathrm{\mu m}$. Our retrievals have a maximum of 14 free parameters, with the parameters and priors summarized in Table~\ref{tab:priors}. We evaluate the parameter space using \texttt{PyMultiNest} \citep{Feroz2008,pymultinest} with 4000 live points for each nested sampling run.


\begin{deluxetable}{lcc}
\tablecaption{Retrieval free parameters and priors.}
\tablehead{
\colhead{Parameter} & \colhead{Prior Distribution} & \colhead{Prior range}
}
\startdata
R$_{p, \, \mathrm{ref}}$ ($R_\mathrm{p}$) & uniform & $0.85 - 1.15$ \\
T (K) & uniform & $400 - 2300$ \\
$\log{\mathrm{X}}$ & uniform & -$12$ -- -$1$ \\
$\log{a}$ & uniform & -$4$ -- $8$ \\
$\gamma$ & uniform & -$20$ -- $2$ \\
$\log{P_\mathrm{cloud}}$ & uniform & -$7$ -- $2$ \\
$\phi_\mathrm{cloud}$ & uniform & $0 - 1$ \\
$\delta_\mathrm{rel}$ (ppm) & uniform & -$1000$ -- $1000$ \\
\enddata
\tablecomments{The reference radius, R$_{p, \, \mathrm{ref}}$, defined at a pressure of 10 bar, spans a prior ranging from 85\%--115\% of the white light radius of WASP-127b (1.311\,$R_J$). The relative offset $\delta_\mathrm{rel}$ is measured in parts per million.}
\label{tab:priors}
\end{deluxetable}

We initially ran retrievals on the G280 data alone, using our base retrieval configuration and models with \ce{Na} and \ce{FeH} excluded. These latter two retrievals allow Bayesian model comparisons to assess the statistical evidence for \ce{Na} and \ce{FeH}. We focus on \ce{Na} and \ce{FeH} as these are the only chemical species with significant absorption features over the UVIS/G280 wavelength range that improve the Bayesian evidence for some model and data combinations. We also conducted the same model comparisons for retrievals combining the G280, \textit{Spitzer}, and subsets of the HST IR observations (G141 and/or STIS). We further consider the influence of a fixed offset between the G280 and IR data (adopting an offset of 195\,ppm for G280). Finally, for comparison with \citet{Spake2021}, we also include a retrieval without our G280 data (only G141, STIS, and \textit{Spitzer}). Our full suite of retrieval models is summarized in Table~\ref{tab:retrieval_model}. We also report in Table~\ref{tab:retrieval_model_stats} statistical metrics for the goodness of fit for all our retrieval models, including the log Bayesian evidence, $\ln \mathcal{Z}$, the Bayesian Information Criterion (BIC), the reduced chi-squared, $\chi^2_{\nu}$, the $\chi^2$ cumulative distribution function (CDF) \citep{wilson2021chisqcdf}, Bayes factors, $\mathcal{B}$, and chemical species detection significances.


\begin{deluxetable*}{lcccccccc}
\tablecaption{\texttt{POSEIDON} retrievals applied to WASP-127b's transmission spectrum.}
\tablehead{
\colhead{\hspace{-1.5cm} Model + Data Combination} & \colhead{Chemical Species} & \colhead{G280 Offset}
}
\startdata \noalign{\vskip 1pt}
\textbf{G280} & & & & \\ \noalign{\vskip 1pt}
\hspace{2pt} Model A & Na + K + \ce{H2O} & --- \\
\hspace{2pt} Model B & K + \ce{H2O} & --- \\
\hspace{2pt} Model C & Na + K + \ce{H2O} + FeH & --- \\ \noalign{\vskip 1pt}
\hline \noalign{\vskip 1pt}
\textbf{G280 + G141 + STIS + \textit{Spitzer}} & & & & & \\ \noalign{\vskip 1pt}
\hspace{2pt} Model D & Na + K + \ce{H2O} + CO + \ce{CH4} + \ce{CO2} & V \\
\hspace{2pt} Model E & K + \ce{H2O} + CO + \ce{CH4} + \ce{CO2} & V \\
\hspace{2pt} Model F & Na + K + \ce{H2O} + CO + \ce{CH4} + \ce{CO2} + FeH & V \\ \noalign{\vskip 1pt}
\hline \noalign{\vskip 1pt}
\textbf{G280 + G141 + \textit{Spitzer}} & & & & & \\ \noalign{\vskip 1pt}
\hspace{2pt} Model G & Na + K + \ce{H2O} + CO + \ce{CH4} + \ce{CO2} & V \\
\hspace{2pt} Model H & K + \ce{H2O} + CO + \ce{CH4} + \ce{CO2} & V \\
\hspace{2pt} Model I & Na + K + \ce{H2O} + CO + \ce{CH4} + \ce{CO2} + FeH & V \\
\hspace{2pt} Model J & Na + K + \ce{H2O} + CO + \ce{CH4} + \ce{CO2} & F (195\,ppm) \\
\hspace{2pt} Model K & K + \ce{H2O} + CO + \ce{CH4} + \ce{CO2} & F (195\,ppm) \\
\hspace{2pt} Model L & Na + K + \ce{H2O} + CO + \ce{CH4} + \ce{CO2} + FeH & F (195\,ppm) \\ \noalign{\vskip 1pt}
\hline \noalign{\vskip 1pt}
\textbf{G141 + STIS + \textit{Spitzer}} & & & & & \\ \noalign{\vskip 1pt}
\hspace{2pt} Model M &  Na + K + \ce{H2O} + CO + \ce{CH4} + \ce{CO2} + \ce{FeH} & --- \\
\enddata
\tablecomments{``STIS'' refers to the combination of G430L and G750L observations. All models contain additional parameters for cloud opacity ($\log a$, $\gamma$, $\log P_\mathrm{cloud}$, and $\phi_\mathrm{cloud}$) and an isothermal temperature profile. Retrievals where the UVIS/G280 data has a free offset are denoted by a ``V'' (Variable offset). Retrievals where the offset is fixed to the median retrieved value from model D (195\,ppm) are denoted with an ``F'' (Fixed).}
\label{tab:retrieval_model}
\end{deluxetable*}

\begin{figure*}
    \centering
    \includegraphics[width=0.9\textwidth]{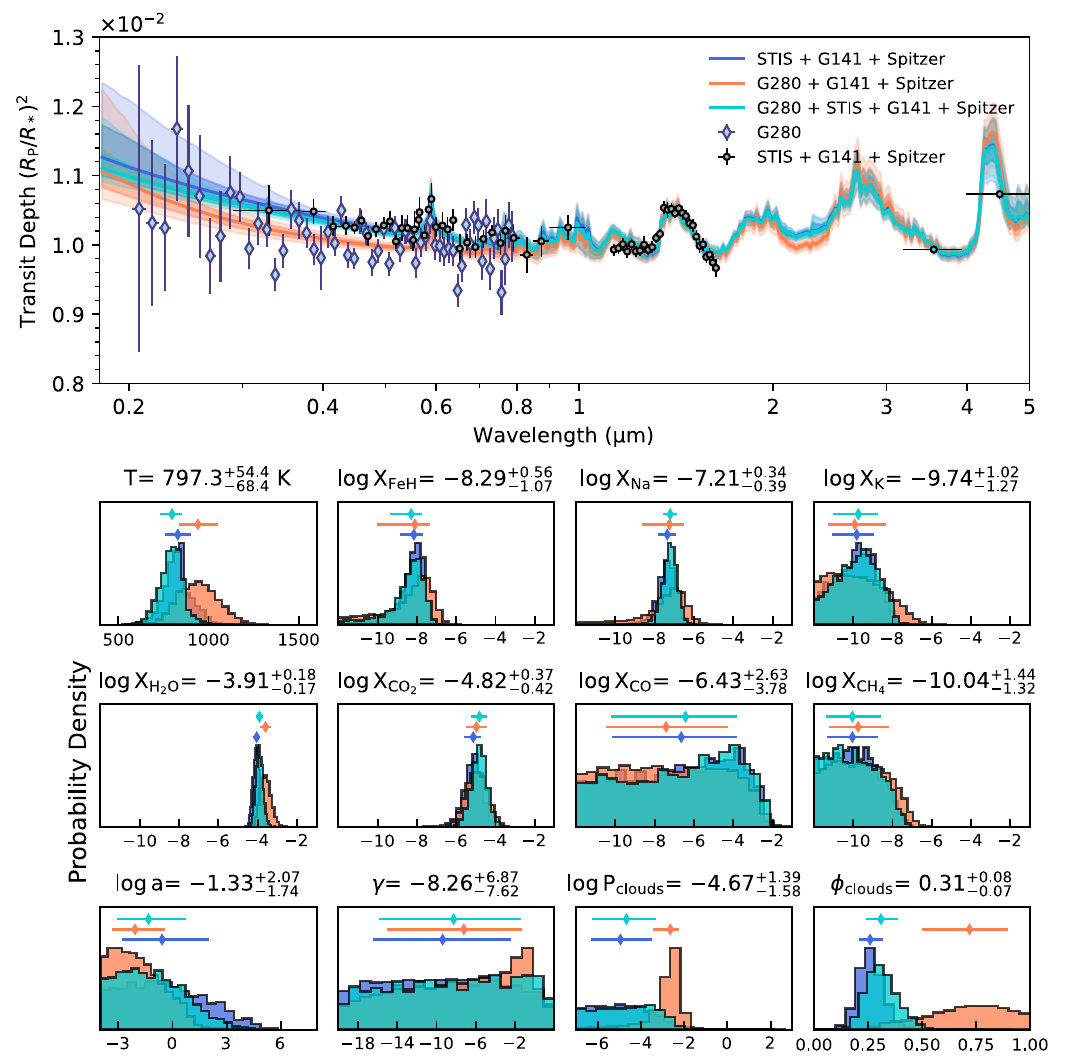}
    \caption{Retrieved \texttt{POSEIDON}  transmission spectra and posterior distributions of WASP-127b. Top: retrieved transmission spectra for three dataset combinations: (i) the combined dataset from G280 + STIS + WFC3/G141 + \textit{Spitzer} (model F, turquoise); (ii) G280 + WFC3/G141 + \textit{Spitzer} (model L, orange); and (iii) STIS + WFC3/G141 + \textit{Spitzer} (model M, blue). We note that Model F allows for a variable offset between the UVIS/G280 and \citet{Spake2021} data (STIS + WFC3/G141 + \textit{Spitzer}), where a median G280 offset of 195\,ppm is found. This offset is applied as a fixed value for model L between the UVIS/G280 and IR data. Shaded regions mark the 1 and 2$\sigma$ confidence intervals and errorbars give the retrieved median parameter values and their respective 1$\sigma$ errors. Retrieved median values and 1$\sigma$ errors are presented for model F.}
    \label{fig:poseidon}
\end{figure*}

\begin{deluxetable*}{lccccccccc}
\tablecaption{Bayesian model comparisons and model fit statistics.}
\tablehead{
\colhead{Model} & \colhead{DOF} & \colhead{$\ln \mathcal{Z}$} & \colhead{BIC} & \colhead{$\chi^2_\nu$} & \colhead{$\chi^2$ CDF (\%)} & \colhead{$\mathcal{B}$} & \colhead{Detection Significance}
}
\startdata
A & 50 & 384.15 & 104.19 & 1.35 & 95.00 & --- & ---  \\
B & 51 & 384.18 & 101.47 & 1.35 & 95.15 & 0.97 & [\ce{Na}] Non-detection \\
C & 49 & 384.27 & 107.91 & 1.37 & 95.64 & 1.13 & [\ce{FeH}] Non-detection \\
\hline
D & 111 & 858.59 & 203.63 & 1.27 & 97.12 & --- & --- \\
E & 112 & 851.85 & 216.88 & 1.42 & 99.77 & 846 & [\ce{Na}] 4.1$\sigma$ \\
F & 110 & 859.54 & 204.98 & 1.25 & 96.10 & 2.49 & [\ce{FeH}] 2.0$\sigma$ \\
\hline 
G & 76 & 604.99 & 163.23 & 1.38 & 98.42 & --- & --- \\
H & 77 & 604.74 & 160.12 & 1.38 & 98.48 & 1.28 & [\ce{Na}] 1.4$\sigma$ \\
I & 75 & 606.15 & 162.59 & 1.33 & 97.04 & 3.19 & [\ce{FeH}] 2.1$\sigma$ \\
J & 77 & 604.60 & 156.27 & 1.33 & 97.19 & --- & --- \\
K & 78 & 603.67 & 153.90 & 1.34 & 97.58 & 2.53 & [\ce{Na}] 2.0$\sigma$ \\
L & 76 & 604.86 & 160.19 & 1.34 & 97.44 & 1.30 & [\ce{FeH}] 1.4$\sigma$  \\
\hline
M & 53 & 473.44 & 114.22 & 1.21 & 85.93 & --- & --- \\
\enddata
\tablecomments{``DOF'' is the number of degrees of freedom ($N_{\rm{data}} - N_{\rm{params}}$), $\mathcal{Z}$ is the Bayesian evidence, ``BIC'' is the Bayesian Information Criterion, $\chi^2_\nu$ is the reduced chi-squared, $\chi^2$ CDF is the chi-squared Cumulative Distribution Function, and $\mathcal{B}$ is the Bayes factor between a model including Na (Models B, E, H, K) or FeH (Models C, F, I, L) and a nested model without Na or FeH.}
\label{tab:retrieval_model_stats}
\end{deluxetable*}

\subsubsection{Retrieval Results}

Our initial retrievals using only the UVIS/G280 dataset resulted in a non-detection of a scattering slope or any significant chemical absorption features. Since the UVIS data alone is essentially featureless, the retrieved temperature has only a 2$\sigma$ upper limit of 849\,K. We therefore focus our subsequent retrieval investigations on joint retrievals between the UVIS/G280 data and the archival HST and \textit{Spitzer} data from \citet{Spake2021} to better constrain WASP-127b's atmospheric properties. We report the median atmospheric parameter values and 1$\sigma$ confidence intervals for our suite of retrievals in Table~\ref{tab:retrieval_appendix}, presented in Appendix~\ref{appendix:poseidon_retrievals}.




Our strongest constraints on WASP-127b's atmosphere arise from jointly retrieving our UVIS/G280 data, alongside HST/STIS, HST/G141, and \textit{Spitzer} photometry. The best-fitting model (model F) results in bounded constraints on H$_2$O ($-3.91^{+0.18}_{-0.17}$~dex), CO$_2$ ($-4.82^{+0.37}_{-0.42}$~dex), Na ($-7.21^{+0.34}_{-0.39}$~dex), and FeH ($-8.29^{-0.56}_{-1.07}$~dex). We detect Na at 4.1$\sigma$ but find only marginal evidence of FeH (2.0$\sigma$). These values are similar to those presented in \citet{Chen2018} and \citet{Spake2021}, respectively. We do not detect K, \ce{CH4}, or CO. We show the retrieved transmission spectrum and posterior distributions from this model in Figure~\ref{fig:poseidon} (turquoise). Our \ce{H2O} and \ce{CO2} constraints imply a slightly sub-solar (0.2$\times$-0.5$\times$) metallicity from \ce{H2O} and a slightly super-solar (15$\times$ to 100$\times$) metallicity from \ce{CO2}, assuming a solar C/O ratio. These values are in agreement with the retrievals performed in \citet{Spake2021}), as well as abundances derived for similar temperature planets \citep[e.g.,][]{Grant2023}. Further infrared data as would be possible from JWST are needed to constrain the full inventory of oxygen- and carbon-bearing species and thus the atmospheric metallicity.


We investigated the robustness of the Na detection via additional retrievals with subsets of the data. In removing the G280 data (Model M), we see the \ce{Na} abundance and constraints remain consistent (Figure~\ref{fig:poseidon}, blue distributions). By instead removing the STIS data and using a fixed G280 offset (model L), the retrieved \ce{Na} abundance remains in agreement with model F, although the abundance constraint is weaker. When G280 is the only optical data, we find a reduced detection significance for \ce{Na} (2.0$\sigma$ with a fixed offset, 1.4$\sigma$ with a variable offset between the G280 and IR data). We note that \citet{fairman2024optical} found a similar \ce{Na} detection significance of 2.9$\sigma$ from a \texttt{POSEIDON} retrieval on the HST/STIS + G141 + \textit{Spitzer} data (but without FeH opacity). Therefore, although neither the STIS nor UVIS/G280 data can confidently detect \ce{Na} independently, their combined information provides a stronger detection (4.1$\sigma$).

The inferred FeH abundance from our retrievals ($\log \ce{FeH} = -8.29^{+0.56}_{-1.07}$) is potentially consistent with expectations from chemical models. Equilibrium chemistry predicts \ce{FeH} mixing ratios between $10^{-10}$ and $10^{-8}$ dex in exoplanet atmospheres with equilibrium temperatures $>$ 1000 K at deep pressures of 1 bar up to millibar pressures \citep{woitke2018chemistry}. The \ce{FeH} abundance drops off significantly with altitude due to the probable sequestration of Fe into clouds  \citep{visscher2010chemistry}. Our best-fit cloudy and cloud-free \texttt{PICASO+Virga} forward models and the retrieved values from \texttt{POSEIDON} agree on the FeH abundance for WASP-127b to within 1$\sigma$. Vertical mixing may contribute to the observed abundance of FeH in WASP-127b’s atmosphere, but is not required given current abundance constraints.
Evidence of metal hydrides has been reported in giant exoplanet atmospheres across a range of equilibrium temperatures \citep[e.g.,][]{evans2018wasp121b, macdonaldmadhu2019hatp26b, sotzen2020wasp127b, Skaf2020, Flagg2023}, including for WASP-127b \citet{Skaf2020,Spake2021}. \citet{Skaf2020} reports from WFC3/IR G141 data alone a higher abundance of \ce{FeH} ($\log \ce{FeH} \sim -5$),  while \citet{Spake2021} finds from G141 + STIS + \textit{Spitzer} data a consistent FeH abundance with our results ($\log \ce{FeH} \sim -7.5$). We stress that the highest FeH detection significance we find is only 2.1$\sigma$ (Model I). Therefore, we conclude that our new G280 data do not provide any additional evidence for FeH in WASP-127b's atmosphere, with the weak inference for FeH driven instead by the WFC3/IR G141 data.





\begin{figure*}
    \centering
    \includegraphics[width=\textwidth]{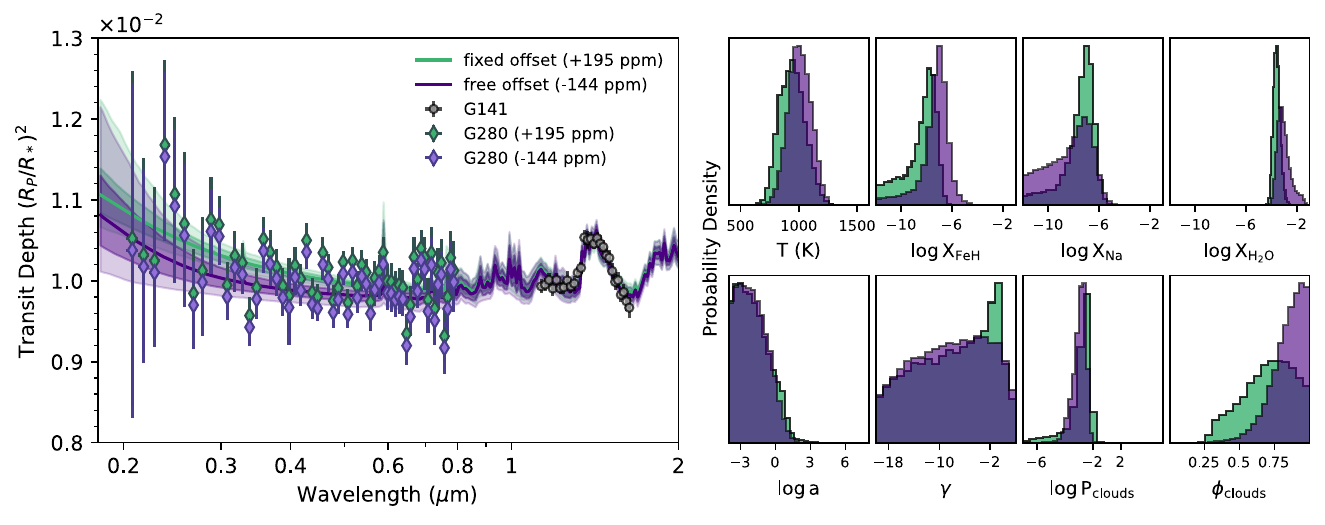}
    \caption{A comparison of the retrieved spectra and selected posterior distributions in allowing a fixed offset at +195ppm (model L, green) to a variable offset (model I, purple) between the UVIS/G280 and IR (WFC3/G141 + \textit{Spitzer}) datasets.
    Shaded regions mark the 1 and 2$\sigma$ confidence intervals}
    \label{fig:poseidon_offsets}
\end{figure*}

Our retrievals favor the presence of an optically thick high-altitude cloud deck. Using the combined data (Model F: G280 + STIS + G141 + \textit{Spitzer}), we find high-altitude patchy clouds with a terminator coverage fraction of $0.31^{+0.08}_{-0.07}$. When fitting for a homogeneous cloud coverage, the probability distribution for the reference radius becomes low, tending towards the lower end of the prior range. This in turn results in an unrealistically tight constraint on gamma which would represent no enhancement over Rayleigh scattering. When allowing for patchy cloud coverage, the reference radius value recovers a Gaussian distribution and we find gamma to be unconstrained at$-8.24^{+6.93}_{-7.58}$, which is more representative of our observational uncertainties where scattering is dominant in the atmosphere ($<$0.5 microns).
However, the inference of patchy clouds depends crucially on the STIS data (0.3--0.9 microns), where a patchy cloud opacity is needed to fit both the flat spectrum and the more prominent Na feature.
When we remove the STIS data and retrieve only the G280 and IR data with a fixed offset (model L), the model lies more centrally through the scatter in the G280 data. This lowers and flattens the model transit depths between 0.4 and 0.8 $\mathrm{\mu m}$, resulting in a uniform cloud fraction with a median retrieved cloud pressure deeper in the atmosphere ($\log P_{\rm{cloud}} \sim 10^{-2.5}$\,bar). Without the STIS data, the retrieved median temperature and H$_2$O abundance increases to compensate for the change from patchy to uniform clouds (see Figure~\ref{fig:poseidon}). These changes are small within the uncertainties of the retrieved parameters, demonstrating that the inferred atmospheric composition of WASP-127b is robust to these permutations of the included optical data. Finally, we show in Figure~\ref{fig:poseidon_offsets} that considering a variable G280 offset instead of a fixed offset (Model I vs. Model L) increases the preference for uniform terminator cloud coverage.

Contrasting with our inference of clouds, we find a non-detection of scattering hazes. This is shown in the posterior distributions in Figure~\ref{fig:poseidon} by the scattering slope parameter, $\gamma$, being unconstrained, and $\log{a}$ having only an upper limit. The non-detection of hazes arises from the small slope at optical wavelengths, which can be explained solely from the contribution of H$_2$ Rayleigh scattering from the clear terminator sector in the patchy cloud model. We find the same absence of aerosol scattering when the G280 data is absent (see Figure~\ref{fig:poseidon}), as the relatively flat UVIS/G280 spectrum is consistent with the STIS observations. We note that our G280 data are inconsistent with the sharp rise in optical transit depths seen in the ground-based observations of WASP-127b in \citet{Palle2017} and \citet{Chen2018}. Therefore, our retrievals indicate an optically thick cloud deck but a lack of small scattering aerosols in WASP-127b's atmosphere.


\section{Discussion} \label{sec:disc}

\subsection{Interpretation of WASP-127b's G280 Spectrum}

We presented \texttt{PICASO} forward models and \texttt{POSEIDON} retrievals in Section~\ref{sec:retrieval} which combined UVIS/G280 observations with previous optical and IR spectra. Both interpretation methods suggest an atmosphere with at least partial cloud cover. The forward model suggests a denser cloud deck below millibar levels and a second, more diffuse cloud deck that extends from around a millibar up to around a microbar; the retrieval supports evidence for patchy clouds.

Theory predicts that at $T_{eq}\sim1400$~K, silicate clouds begin to sink from the observable atmosphere and can be found only at high pressure layers deep in the atmosphere, if they are found at all, while manganese sulfide clouds begin to condense in the upper atmosphere \citep[e.g.,][]{visscher2010chemistry,Parm2016}. This is roughly in line with the results of our \texttt{PICASO+Virga} models, where we observe that the millibar-microbar cloud layer is composed of minor amounts of manganese sulfide clouds along with silicate clouds. The deep, denser cloud deck is entirely silicate. Future studies should pursue these claims with JWST MIRI, which can resolve absorption features unique to silicate species \citep[e.g.,][]{Grant2023}, though MnS lacks distinctive mid-infrared absorption \citep{WakefordSing2015}. Our best-fit \texttt{Virga} models, in particular, suggest strong silicate opacity for WASP-127b (middle panel, Figure~\ref{fig:models}), which JWST MIRI could observe. Moreover, \ce{SiO2} clouds could also contribute to the silicate cloud layer \citep[e.g.,][]{Grant2023}, and MIRI observations could determine the specific identity of silicate clouds which is beyond the capability of Hubble.

SNR limits the ability of the added G280 data to strongly constrain cloud properties via the scattering gradient and Rayleigh enhancement factor from \texttt{POSEIDON} or mixing and sedimentation parameters from \texttt{Virga}. Nonetheless, the UVIS/G280 spectrum is consistent with a shallow slope and the overall flatness of the spectrum suggests a Rayleigh scattering haze flattened by patchy cloud cover. Small particle sizes under 1 micron are favored by the UVIS/G280 spectrum combined with the STIS and G141 data. Weak evidence for sodium is retrieved, but no evidence is found for potassium or lithium, consistent with previous optical and IR studies of WASP-127b with HST \citep{Spake2021}. While the overall spectrum shape is flat or slightly inclined, modest evidence of H$_2$O and CO$_2$ absorption is recovered as is weak evidence of sodium absorption, consistent with partial cloud coverage that mutes, but does not fully suppress, all absorption features. 

Our UVIS/G280 data can provide only upper bounds on the amount of CH$_4$ and CO, as to be expected since neither species has any strong absorption features in the UVIS/G280 wavelength range; JWST observations should be carried out to constrain the CH$_4$ abundance and break the CO/CO$_2$ degeneracy for WASP-127b.

\subsection{Direct Comparison to HST WFC3/IR and STIS Spectroscopy}
As WASP-127b has been studied by HST spectroscopy before, our work provides an important benchmark of the performance of the UVIS/G280 as compared to other HST instruments. We carried out an extraction emulating the work of \cite{Spake2021}, benchmarking the observing potential of the underutilized UVIS/G280 against the more standard STIS G430L+G750L combination used by other studies. The UVIS/G280 throughput in the near- to mid-UV ranges from around 10\% to 30\%, considerably higher than the $<$5\% throughput of the STIS G430L in the near-UV. \cite{Spake2021} utilized the STIS G430L and achieved a precision of about 400~ppm in a spectral bin spanning 289.8~nm to 370.0~nm; in this same wavelength bin and following identical fitting procedures, our UVIS/G280 observations achieved a precision of 66~ppm, a factor of 6 times more precise than what the STIS G430L achieved. However, we note the UVIS/G280 spectrum extracted here shows notably more scatter than that obtained by STIS; data points in the STIS spectrum lie closer to a smooth line while the UVIS/G280 spectrum shows a notable ``zigzag'' pattern. Unusual sky background values have been observed in other UVIS/G280 transmission observations \citep{Wakeford2020} and are believed to be a UVIS/G280-specific systematic signal that is not yet understood; the zigzag seen here likely originates from the same unusual background behavior. Removal of this suspected systematic is expected to remove the zigzag pattern without significantly changing the results of this work. The superior throughput of the UVIS/G280 in near-UV wavelengths thus has the potential to place stronger constraints of a hot Jupiter's UV spectrum than can be achieved with STIS grisms, and further work should be carried out to better characterize and treat its instrument systematics to unlock its full potential.

\subsection{Other studies of WASP-127b}
Our HST UVIS/G280 spectrum, like the HST STIS and IR spectra presented in \citet{Spake2021}, shows no evidence of lithium or potassium. Previous studies using ground-based instruments reported detections of lithium and potassium to $3.4\sigma$ and $5.0\sigma$, respectively \citep{Chen2018}. Telluric O$_2$ absorption features can complicate ground-based observations of exoplanetary potassium \citep[e.g.,][]{Gibson2017}; we therefore suspect that previous observations of lithium and potassium may instead be telluric contamination. Our forward models favor sub-micron particle sizes as were favored by \citet{Chen2018}; however, with the increased precision achievable with HST, we find stronger evidence for a sub-Rayleigh scattering and shallow gradient, consistent with the sub-Rayleigh slope found by \citet{Spake2021}. As stated in Section~\ref{sec:analysis}, the larger uncertainties in the spectra of \citet{Palle2017} and \citet{Chen2018} are consistent with our results, even as the previously-reported super-Rayleigh slope has disappeared.

Recent ground-based high-resolution cross-correlation studies have also detected H$_2$O \citep{Boucher2023} and both H$_2$O and CO in WASP-127b \citep{Kanumalla2024,Nortmann2024}. Observations with these high resolution spectrographs have resulted in contradictory interpretations of the carbon content and metallicity of WASP-127b's atmosphere. The Spectro-Polarimètre InfraRouge (SPIRou) spectrograph on the Canada-France-Hawaii telescope found substellar C/O and only upper limits of CO \citep{Boucher2023}, while observations with the  Immersion GRating
INfrared Spectrometer (IGRINS) on Gemini South and the CRyogenic InfraRed Echelle Spectrograph (CRIRES$^+$) of the European Southern Observatory both have higher SNR and were able to recover CO constraints for the planet \citep{Kanumalla2024,Nortmann2024}, indicating sub-stellar to stellar C/O. 

Similarly, the metallicity constraints enabled by these high-resolution cross-correlation studies vary from solar \citep{Nortmann2024} to significantly metal-rich ($\sim$40$\times$ solar; \cite{Boucher2023,Kanumalla2024}). These varied interpretations are in line with our retrievals on the combined G280 + STIS + G141 + \textit{Spitzer} data, where both our study and the cross-correlation interpretations rely on the \textit{Spitzer} photometry for contributions from CO$_2$ as the most powerful tracer of atmospheric metallicity. Both \citet{Nortmann2024} and \citet{Boucher2023} find \ce{H2O} abundances in agreement to 1$\sigma$ with those we detect here, while \citet{Kanumalla2024} retrieves a \ce{H2O} VMR two orders of magnitude higher than our 1$\sigma$ bound.

These high-resolution cross-correlation (HRCC) studies all also find evidence for a substantial grey cloud deck which ranges from 10s of millibar \citep{Boucher2023} to 1--0.1 millibar \citep{Kanumalla2024,Nortmann2024}, in agreement with our space-based data interpretation. Interestingly, the three cross-correlation studies also all find evidence of significant blue-shifted \citep{Boucher2023,Kanumalla2024,Nortmann2024} and red-shifted \citep{Nortmann2024} signal peaks, indicative of a super-sonic equatorial jet and significant day-night temperature contrasts. This inferred atmospheric profile from HRCC, combined with our cloud parameter constraints, strongly supports JWST follow-up studies of this intriguing world to better understand its temperature structure, chemistry, and clouds. 

\section{Conclusions} \label{sec:outro}
We present the UVIS transmission spectrum of WASP-127b taken through the HST WFC3/UVIS G280 as part of the HUSTLE program. Our primary spectrum spans 200~nm to 800~nm, achieving a broad-band precision of 91~ppm and a median precision of $\sim240$~ppm in 10~nm-wide bins. A supplementary reduction using an independent pipeline was statistically consistent with the results of our primary reduction, supporting the claim by \citet{Wakeford2020} that G280 observations are consistent enough to provide meaningful results independent of particular reduction and analysis techniques.

Forward models show a strong preference for the presence of clouds, and while not strongly constrained, all models favor clouds of sub-micron particles at relatively high altitudes. Attempts to fit clear atmospheric models to the data with forward modeling yield fits of poor quality requiring very supersolar metallicities, disfavoring the clear sky interpretation of WASP-127b’s transmission spectrum. Retrievals recover modest evidence for Na absorption with an overall detection confidence of $4.1\sigma$ when G280 is combined with archival space-based transmission spectra. Any observed signature of Na absorption suggests that the cloud coverage cannot be total. Our retrievals favor a partial cloud coverage fraction of $\phi\sim0.31$. Evidence for FeH, previously inferred from G141 data by \citet{Skaf2020} ($4.0\sigma$) and \citet{Spake2021} ($2.0\sigma$), is not strengthened with the inclusion of G280 data.

Extending the spectral coverage down to mid-UV wavelengths shows that the scattering slope, while weakly constrained here, is sub-Rayleigh in nature, consistent with claims by \citet{Spake2021} and contradictory to results obtained from the ground \citep{Palle2017,Chen2018}. The small slope can be accounted for with H$_2$ Rayleigh scattering from the clear component of the terminator, with an opaque cloud deck elsewhere. The equilibrium temperature of WASP-127b is in the correct range to be at a transition point in aerosols where silicate clouds subside and manganese sulfide clouds begin to condense in the upper atmosphere \citep[e.g.][]{Parm2016}. We recommend future studies of WASP-127b complement this work through JWST observations in the near- and mid-IR, where absorption signatures from silicates can be detected, confirming the partial clouds found here and constraining their composition.

Our study demonstrates the potential for the UVIS/G280 to be a more efficient observing mode for UVIS spectroscopy than what is typically achieved using HST STIS. In comparison to previous studies of WASP-127b through STIS observing modes \citep{Spake2021}, our UVIS/G280 observations achieve a near-UV (289.8~nm to 370.0~nm) precision of 66~ppm, about $6\times$ greater precision than was achieved by the standard STIS G430L+G750L observing mode combination, albeit with higher scatter for which we recommend further study of the instrument's systematics to resolve. We achieve this precision in just one transit observation in comparison the two-transit minimum that is imposed by STIS spectroscopy. The HST WFC3/UVIS G280’s spectral range of 200~nm to 800~nm provides a much deeper probe into the UV than can be achieved with STIS G430L+G750L, while providing enough overlap with JWST (lower limit of 600~nm) to allow for offsets to be applied to match G280 observations to complementary JWST observations. We therefore strongly recommend the use of UVIS/G280 spectroscopy as a powerful UV complement to JWST IR spectroscopy, opening a profound new visual landscape into exoplanet atmospheres from the mid-UV to the mid-IR.

\begin{acknowledgments}
This research is based on observations made with the NASA/ESA \textit{Hubble Space Telescope} obtained from the Space Telescope Science Institute, which is operated by the Association of Universities for Research in Astronomy, Inc., under NASA contract NAS 5–26555. These observations are associated with program HST-GO 17183, PI: H.R. Wakeford. This research has made use of the NASA Exoplanet Archive, which is operated by the California Institute of Technology, under contract with the National Aeronautics and Space Administration under the Exoplanet Exploration Program.
V.A.B., N.K.L., S.E.M., M.L-M., M.S.M, N.E.B., acknowledge support for program number HST-GO-17183 provided through a grant from the STScI under NASA contract NAS5-26555.
C.E.F. acknowledges funding from the University of Bristol School of Physics PhD Scholarship Fund. 
C.G. acknowledges funding from La Caixa Fellowship and the Agency for Management of University and Research Grants from the Government of Catalonia (FI AGAUR).
H.R.W. and D.G were funded by UK Research and Innovation (UKRI) framework under the UK government’s Horizon Europe funding guarantee for an ERC Starter Grant [grant number EP/Y006313/1]. 
L.A acknowledges funding from the UKRI STFC Consolidated Grant ST/V000454/1. 
J.K.B. is supported by an STFC Ernest Rutherford Fellowship, grant ST/T004479/1.
R.J.M. is supported by NASA through the NASA Hubble Fellowship grant HST-HF2-51513.001, awarded by the Space Telescope Science Institute, which is operated by the Association of Universities for Research in Astronomy, Inc., for NASA, under contract NAS 5-26555. We thank the anonymous reviewer for their swift and insightful feedback.

\end{acknowledgments}

%
All the HST data used in this paper can be found in MAST: \dataset[10.17909/yrft-8f93]{http://dx.doi.org/10.17909/yrft-8f93}
\vspace{5mm}
\facility{HST(WFC3)}


\software{astropy \citep{2013A&A...558A..33A,2018AJ....156..123A},  
          scipy \citep{Scipy}, 
          ExoTiC-LD \citep{Exotic2022},
          ExoTiC-ISM \citep{Exoticism},
          batman \citep{Batman},
          emcee \citep{Emcee}
          }



\newpage
\appendix

\section{HST WFC3/UVIS G280 Transmission Spectrum of WASP-127b}

\begin{longtable*}[H]{l|l|l|l|l|l}
    \hline
    \hline
    Wavelength [nm] & Transit depth [$\%$] & $u_1$ & $u_2$ & $u_3$ & $u_4$ \\
    \hline
    $207.5$ & $1.052\pm0.207$ & $-0.247$ & $2.203$ & $-4.652$ & $3.654$ \\
    $217.5$ & $1.032\pm0.119$ & $0.029$ & $1.145$ & $-2.816$ & $2.614$ \\
    $227.5$ & $1.025\pm0.092$ & $0.326$ & $1.033$ & $-2.318$ & $1.942$ \\
    $237.5$ & $1.168\pm0.105$ & $0.399$ & $0.213$ & $-1.153$ & $1.528$ \\
    $247.5$ & $1.107\pm0.095$ & $0.554$ & $-0.579$ & $0.619$ & $0.387$ \\
    $257.5$ & $1.070\pm0.089$ & $0.323$ & $-0.850$ & $1.942$ & $-0.459$ \\
    $267.5$ & $0.984\pm0.055$ & $0.751$ & $-1.008$ & $1.275$ & $-0.035$ \\
    $277.5$ & $1.013\pm0.066$ & $0.466$ & $-0.930$ & $1.879$ & $-0.455$ \\
    $287.5$ & $1.075\pm0.053$ & $0.355$ & $-0.799$ & $2.089$ & $-0.693$ \\
    $297.5$ & $1.069\pm0.036$ & $0.312$ & $-0.582$ & $1.896$ & $-0.683$ \\
    $307.5$ & $0.995\pm0.029$ & $0.313$ & $-0.393$ & $1.595$ & $-0.584$ \\
    $317.5$ & $1.031\pm0.030$ & $0.230$ & $-0.096$ & $1.374$ & $-0.584$ \\
    $327.5$ & $1.022\pm0.025$ & $0.457$ & $-0.670$ & $1.845$ & $-0.698$ \\
    $337.5$ & $0.957\pm0.024$ & $0.410$ & $-0.558$ & $1.876$ & $-0.801$ \\
    $347.5$ & $0.992\pm0.025$ & $0.491$ & $-0.704$ & $2.013$ & $-0.866$ \\
    $357.5$ & $1.051\pm0.028$ & $0.339$ & $-0.042$ & $1.220$ & $-0.607$ \\
    $367.5$ & $1.035\pm0.027$ & $0.584$ & $-0.993$ & $2.285$ & $-0.943$ \\
    $377.5$ & $1.017\pm0.022$ & $0.758$ & $-1.331$ & $2.285$ & $-0.774$ \\
    $387.5$ & $0.994\pm0.022$ & $0.642$ & $-0.826$ & $1.799$ & $-0.696$ \\
    $397.5$ & $0.981\pm0.046$ & $0.373$ & $-0.153$ & $1.263$ & $-0.578$ \\
    $407.5$ & $1.019\pm0.023$ & $0.421$ & $-0.337$ & $1.472$ & $-0.648$ \\
    $417.5$ & $1.002\pm0.019$ & $0.555$ & $-0.562$ & $1.481$ & $-0.585$ \\
    $427.5$ & $1.050\pm0.021$ & $0.522$ & $-0.506$ & $1.529$ & $-0.659$ \\
    $437.5$ & $0.985\pm0.018$ & $0.374$ & $0.049$ & $0.929$ & $-0.472$ \\
    $447.5$ & $0.980\pm0.015$ & $0.343$ & $0.217$ & $0.711$ & $-0.399$ \\
    $457.5$ & $1.035\pm0.019$ & $0.364$ & $0.167$ & $0.760$ & $-0.424$ \\
    $467.5$ & $1.017\pm0.016$ & $0.359$ & $0.226$ & $0.663$ & $-0.390$ \\
    $477.5$ & $0.975\pm0.019$ & $0.433$ & $0.064$ & $0.819$ & $-0.467$ \\
    $487.5$ & $0.991\pm0.017$ & $0.432$ & $0.069$ & $0.773$ & $-0.428$ \\
    $497.5$ & $1.023\pm0.014$ & $0.449$ & $0.018$ & $0.829$ & $-0.453$ \\
    $507.5$ & $0.973\pm0.017$ & $0.540$ & $-0.264$ & $1.099$ & $-0.543$ \\
    $517.5$ & $1.025\pm0.017$ & $0.504$ & $-0.082$ & $0.863$ & $-0.457$ \\
    $527.5$ & $0.993\pm0.018$ & $0.501$ & $-0.058$ & $0.827$ & $-0.449$ \\
    $537.5$ & $1.020\pm0.018$ & $0.506$ & $-0.063$ & $0.823$ & $-0.449$ \\
    $547.5$ & $1.011\pm0.017$ & $0.520$ & $-0.083$ & $0.815$ & $-0.443$ \\
    $557.5$ & $0.974\pm0.021$ & $0.535$ & $-0.119$ & $0.837$ & $-0.450$ \\
    $567.5$ & $1.003\pm0.020$ & $0.524$ & $-0.073$ & $0.775$ & $-0.428$ \\
    $577.5$ & $1.012\pm0.020$ & $0.539$ & $-0.112$ & $0.798$ & $-0.435$ \\
    $587.5$ & $1.035\pm0.019$ & $0.535$ & $-0.081$ & $0.745$ & $-0.415$ \\
    $597.5$ & $1.000\pm0.021$ & $0.548$ & $-0.114$ & $0.764$ & $-0.419$ \\
    $607.5$ & $0.999\pm0.023$ & $0.592$ & $-0.240$ & $0.880$ & $-0.459$ \\
    $617.5$ & $0.991\pm0.021$ & $0.597$ & $-0.247$ & $0.873$ & $-0.455$ \\
    $627.5$ & $0.995\pm0.021$ & $0.600$ & $-0.262$ & $0.888$ & $-0.465$ \\
    $637.5$ & $0.992\pm0.021$ & $0.613$ & $-0.285$ & $0.890$ & $-0.462$ \\
    $647.5$ & $0.934\pm0.024$ & $0.653$ & $-0.306$ & $0.841$ & $-0.451$ \\
    $657.5$ & $0.970\pm0.024$ & $0.618$ & $-0.309$ & $0.901$ & $-0.466$ \\
    $667.5$ & $1.029\pm0.025$ & $0.619$ & $-0.312$ & $0.886$ & $-0.455$ \\
    $677.5$ & $1.002\pm0.021$ & $0.629$ & $-0.340$ & $0.902$ & $-0.458$ \\
    $687.5$ & $1.040\pm0.023$ & $0.644$ & $-0.386$ & $0.946$ & $-0.475$ \\
    $697.5$ & $1.029\pm0.025$ & $0.651$ & $-0.407$ & $0.957$ & $-0.477$ \\
    $707.5$ & $0.977\pm0.027$ & $0.674$ & $-0.474$ & $1.012$ & $-0.494$ \\
    $717.5$ & $1.034\pm0.033$ & $0.669$ & $-0.461$ & $0.991$ & $-0.486$ \\
    $727.5$ & $0.965\pm0.030$ & $0.675$ & $-0.481$ & $1.001$ & $-0.488$ \\
    $737.5$ & $1.010\pm0.031$ & $0.671$ & $-0.469$ & $0.979$ & $-0.480$ \\
    $747.5$ & $1.022\pm0.029$ & $0.664$ & $-0.446$ & $0.941$ & $-0.463$ \\
    $757.5$ & $0.932\pm0.032$ & $0.672$ & $-0.470$ & $0.957$ & $-0.469$ \\
    $767.5$ & $0.980\pm0.037$ & $0.664$ & $-0.444$ & $0.912$ & $-0.447$ \\
    $777.5$ & $1.028\pm0.034$ & $0.675$ & $-0.479$ & $0.953$ & $-0.465$ \\
    $787.5$ & $1.013\pm0.037$ & $0.691$ & $-0.532$ & $1.005$ & $-0.483$ \\
    \hline
    \hline
    \caption[width=0.4\columnwidth]{Atmospheric transmission spectrum of WASP-127b derived from HST WFC3/UVIS G280 observations and extracted with \Hazelnut. We include the central wavelengths for our bins (all with halfwidth 5~nm) and limb darkening coefficients ($u_i$) for our 4-parameter nonlinear limb darkening law.}
    \label{tab:spectrum}
\end{longtable*}

\section{Retrieved POSEIDON atmospheric parameters}\label{appendix:poseidon_retrievals}
\newpage

\renewcommand{\arraystretch}{1.1}

\begin{deluxetable*}{lcccccccccccccc}
\tablecaption{Median and 1$\sigma$ errors of atmospheric parameters for each POSEIDON retrieval model.}\label{tab:retrieval_appendix}
\tablehead{
\colhead{Model} & \colhead{$\log{X_{Na}}$} & \colhead{$\log{X_{K}}$} & \colhead{$\log{X_{H_2O}}$} & \colhead{$\log{X_{CO_2}}$} & \colhead{$\log{X_{CO}}$} & \colhead{$\log{X_{CH_4}}$} & \colhead{$\log{X_{FeH}}$}
}
\startdata
A & $-8.13^{+1.53}_{-2.14}$ & $-9.50^{+1.95}_{-1.58}$ & $-7.17^{+3.62}_{-3.07}$ & - & - & - & - \\
B & - & $-9.57^{+1.93}_{-1.54}$ & $-6.93^{+3.61}_{-3.30}$ & - & - & - & - \\
C & $-8.05^{+1.42}_{-2.07}$ & $-9.67^{+1.93}_{-1.47}$ & $-6.90^{+3.53}_{-3.18}$ & - & - & - & $-7.75^{+2.43}_{-2.63}$ \\
\hline
D & $-7.28^{+0.37}_{-0.44}$ & $-9.42^{+0.93}_{-1.36}$ & $-3.89^{+0.20}_{-0.17}$ & $-4.75^{+0.39}_{-0.38}$ & $-6.40^{+2.71}_{-3.78}$ & $-10.09^{+1.47}_{-1.28}$ & - \\
E & - & $-9.92^{+0.94}_{-1.22}$ &  $-3.99^{+0.20}_{-0.17}$ & $-4.84^{+0.39}_{-0.42}$ & $-6.09^{+2.43}_{-3.77}$ & $-10.14^{+1.36}_{-1.25}$ & - \\
F & $-7.21^{+0.34}_{-0.39}$ & $-9.74^{+1.02}_{-1.27}$ & $-3.91^{+0.18}_{-0.17}$ & $-4.82^{+0.37}_{-0.42}$ & $-6.43^{+2.63}_{-3.78}$ & $-10.04^{+1.44}_{-1.32}$ & $-8.29^{+0.56}_{-1.07}$ \\
\hline
G & $-8.08^{+1.54}_{-2.46}$ & $-9.69^{+1.77}_{-1.55}$ & $-3.06^{+0.70}_{-0.45}$ & $-4.52^{+0.91}_{-0.69}$ & $-7.37^{+3.23}_{-3.10}$ & $-9.47^{+1.77}_{-1.69}$ & - \\
H & - & $-9.72^{+1.79}_{-1.53}$ & $-3.03^{+0.65}_{-0.45}$ & $-4.48^{+0.85}_{-0.69}$ & $-7.36^{+3.27}_{-3.10}$ & $-9.47^{+1.81}_{-1.71}$ & - \\
I & $-8.06^{+1.38}_{-2.41}$ & $-9.95^{+1.68}_{-1.37}$ & $-3.14^{+0.60}_{-0.39}$ & $-4.70^{+0.84}_{-0.63}$ & $-7.24^{+3.27}_{-3.19}$ & $-9.47^{+1.82}_{-1.69}$ & $-7.17^{+0.70}_{-1.07}$ \\
J & $-7.38^{+0.91}_{-2.07}$ & $-9.74^{+1.67}_{-1.51}$ & $-3.55^{+0.36}_{-0.30}$ & $-4.94^{+0.63}_{-0.59}$ & $-7.41^{+3.15}_{-3.11}$ & $-9.77^{+1.63}_{-1.47}$ & - \\
K & - & $-9.79^{+1.71}_{-1.48}$ & $-3.56^{+0.37}_{-0.30}$ & $-4.91^{+0.63}_{-0.60}$ & $-7.44^{+3.15}_{-3.07}$ & $-9.74^{1.66}_{-1.52}$ & - \\
L & $-7.25^{+0.72}_{-1.41}$ & $-9.92^{+1.61}_{-1.38}$ & $-3.60^{+0.29}_{-0.26}$ & $-4.97^{+0.56}_{-0.55}$ & $-7.42^{+3.17}_{-3.07}$ & $-9.75^{+1.57}_{-1.47}$ & 
$-8.10^{+0.77}_{-1.92}$ \\
\hline
M & $-7.36^{+0.42}_{-0.48}$ & $-9.82^{+0.90}_{-1.26}$ & $-4.05^{+0.18}_{-0.17}$ & $-5.13^{+0.41}_{-0.48}$ & $-6.65^{+2.82}_{-3.54}$ & $-10.04^{+1.32}_{-1.28}$ & $-8.15^{+0.44}_{-0.72}$\\
\hline
\hline
\colhead{Model} & \colhead{$R_\mathrm{P_{ref}}$ (R$_\mathrm{J}$) } & \colhead{$T$ (K)} & \colhead{$\log{a}$} & \colhead{$\gamma$} & \colhead{$\log{P_{cloud}}$} & \colhead{$\bar{\phi}_{cloud}$} & \colhead{$\delta_\mathrm{rel}$ (ppm)} \\
\hline
A & $1.19^{+0.03}_{-0.04}$ & $565.0^{+201.0}_{-113.2}$ & $-0.78^{+2.50}_{-2.01}$ & $-8.48^{+7.51}_{-7.31}$ & $-3.73^{+1.33}_{-1.39}$ & $0.72^{+0.16}_{-0.22}$ & - \\
B & $1.18^{+0.03}_{-0.04}$ & $573.0^{+201.6}_{-118.8}$ & $-0.77^{+2.50}_{-2.02}$ & $-7.94^{+7.07}_{-7.61}$ & $-3.74^{+1.24}_{-1.41}$ & $0.74^{+0.15}_{-0.20}$ & - \\
C & $1.19^{+0.03}_{-0.04}$ & $550.2^{+188.7}_{-103.1}$ & $-0.73^{+2.52}_{-2.04}$ & $-8.12^{+6.92}_{-7.37}$ & $-3.74^{+1.36}_{-1.36}$ & $0.68^{+0.18}_{-0.24}$ & - \\
\hline
D & $1.19^{+0.02}_{-0.02}$ & $770.7^{+59.5}_{-69.9}$ & $-1.59^{+2.01}_{-1.64}$ & $-8.24^{+6.93}_{-7.58}$ & $-4.07^{+1.12}_{ -1.74}$ & $0.35^{+0.08}_{-0.07}$ & $-195.05^{+38.29}_{-40.59}$\\
E & $1.20^{+0.02}_{-0.02}$ & $751.1^{+62.6}_{-73.7}$ & $-1.75^{+1.80}_{-1.49}$ & $-8.26^{+6.93}_{-7.28}$ & $-3.66^{+0.92}_{-1.87}$ & $0.34^{+0.10}_{-0.09}$ & $-190.79^{+37.12}_{ -39.37}$ \\
F & $1.18^{+0.02}_{-0.02}$ & $797.3^{+54.4}_{-68.4}$ & $-1.33^{+2.07}_{-1.74}$ & $-8.26^{+6.87}_{-7.62}$ & $-4.67^{+1.39}_{-1.58}$ & $0.31^{+0.08}_{-0.07}$ & $-190.24^{+39.94}_{-38.61}$ \\
\hline
G & $1.14^{+0.02}_{-0.01}$ & $968.1^{+102.5}_{-93.1}$ & $-2.02^{+1.53}_{-1.32}$ & $-8.28^{+6.38}_{-7.12}$ & $-2.87^{+0.37}_{-0.57}$ & $0.92^{+0.06}_{-0.10}$ & $144.25^{+59.15}_{-58.39}$ \\
H & $1.14^{+0.02}_{-0.01}$ & $965.3^{+99.7}_{-93.2}$ & $-2.05^{+1.51}_{-1.31}$ & $-8.23^{+6.45}_{-7.13}$ & $-2.90^{+0.38}_{-0.53}$ & $0.92^{+0.06}_{-0.10}$ & $154.59^{+55.51}_{-55.37}$\\
I & $1.13^{+0.02}_{-0.01}$ & $1005.6^{+101.8}_{-94.2}$ & $-2.10^{+1.49}_{-1.26}$ & $-8.07^{+6.31}_{-7.18}$ & $-2.87^{+0.36}_{-0.56}$ & $0.87^{+0.09}_{-0.14}$ & $191.85^{+68.56}_{-69.81}$ \\
J & $1.16^{+0.02}_{-0.02}$ & $946.5^{+123.5}_{-117.4}$ & $-2.00^{+2.16}_{-1.37}$ & $-5.82^{+4.88}_{-8.50}$ & $-2.50^{+0.29}_{-0.42}$ & $0.86^{+0.10}_{-0.14}$ & -  \\
K & $1.16^{+0.02}_{-0.02}$ & $941.2^{+122.1}_{-118.4}$ & $-1.93^{+2.36}_{-1.41}$ & $-5.06^{+4.25}_{-9.07}$ & $-2.49^{+0.29}_{-0.42}$ & $0.87^{+0.10}_{-0.16}$ & - \\
L & $1.16^{+0.02}_{-0.02}$ & $941.3^{+113.7}_{-107.6}$ & $-2.08^{+1.69}_{-1.30}$ & $-7.22^{+5.95}_{-7.79}$ & $-2.64^{+0.38}_{-0.78}$ & $0.72^{+0.18}_{-0.22}$ & -\\
\hline
M & $1.18^{+0.02}_{-0.02}$ & $829.5^{+75.0}_{-68.6}$ & $-0.58^{+2.60}_{-2.20}$ & $-9.37^{+6.99}_{-7.10}$ & $-4.95^{+1.46}_{-1.36}$ & $0.26^{+0.06}_{-0.05}$ & - \\
\enddata
\end{deluxetable*}


\bibliography{refs}{}

\begin{thebibliography}{}
\expandafter\ifx\csname natexlab\endcsname\relax\def\natexlab#1{#1}\fi
\providecommand{\url}[1]{\href{#1}{#1}}
\providecommand{\dodoi}[1]{doi:~\href{http://doi.org/#1}{\nolinkurl{#1}}}
\providecommand{\doeprint}[1]{\href{http://ascl.net/#1}{\nolinkurl{http://ascl.net/#1}}}
\providecommand{\doarXiv}[1]{\href{https://arxiv.org/abs/#1}{\nolinkurl{https://arxiv.org/abs/#1}}}

\bibitem[{{Ackerman} \& {Marley}(2001)}]{virga_methods}
{Ackerman}, A.~S., \& {Marley}, M.~S. 2001, \apj, 556, 872, \dodoi{10.1086/321540}

\bibitem[{{Ahrer} {et~al.}(2023){Ahrer}, {Stevenson}, {Mansfield}, {Moran}, {Brande}, {Morello}, {Murray}, {Nikolov}, {Petit dit de la Roche}, {Schlawin}, {Wheatley}, {Zieba}, {Batalha}, {Damiano}, {Goyal}, {Lendl}, {Lothringer}, {Mukherjee}, {Ohno}, {Batalha}, {Battley}, {Bean}, {Beatty}, {Benneke}, {Berta-Thompson}, {Carter}, {Cubillos}, {Daylan}, {Espinoza}, {Gao}, {Gibson}, {Gill}, {Harrington}, {Hu}, {Kreidberg}, {Lewis}, {Line}, {L{\'o}pez-Morales}, {Parmentier}, {Powell}, {Sing}, {Tsai}, {Wakeford}, {Welbanks}, {Alam}, {Alderson}, {Allen}, {Anderson}, {Barstow}, {Bayliss}, {Bell}, {Blecic}, {Bryant}, {Burleigh}, {Carone}, {Casewell}, {Changeat}, {Chubb}, {Crossfield}, {Crouzet}, {Decin}, {D{\'e}sert}, {Feinstein}, {Flagg}, {Fortney}, {Gizis}, {Heng}, {Iro}, {Kempton}, {Kendrew}, {Kirk}, {Knutson}, {Komacek}, {Lagage}, {Leconte}, {Lustig-Yaeger}, {MacDonald}, {Mancini}, {May}, {Mayne}, {Miguel}, {Mikal-Evans}, {Molaverdikhani}, {Palle}, {Piaulet}, {Rackham}, {Redfield}, {Rogers}, {Roy}, {Rustamkulov},
  {Shkolnik}, {Sotzen}, {Taylor}, {Tremblin}, {Tucker}, {Turner}, {de Val-Borro}, {Venot}, \& {Zhang}}]{Ahrer2023}
{Ahrer}, E.-M., {Stevenson}, K.~B., {Mansfield}, M., {et~al.} 2023, \nat, 614, 653, \dodoi{10.1038/s41586-022-05590-4}

\bibitem[{{Alderson} {et~al.}(2023){Alderson}, {Wakeford}, {Alam}, {Batalha}, {Lothringer}, {Adams Redai}, {Barat}, {Brande}, {Damiano}, {Daylan}, {Espinoza}, {Flagg}, {Goyal}, {Grant}, {Hu}, {Inglis}, {Lee}, {Mikal-Evans}, {Ramos-Rosado}, {Roy}, {Wallack}, {Batalha}, {Bean}, {Benneke}, {Berta-Thompson}, {Carter}, {Changeat}, {Col{\'o}n}, {Crossfield}, {D{\'e}sert}, {Foreman-Mackey}, {Gibson}, {Kreidberg}, {Line}, {L{\'o}pez-Morales}, {Molaverdikhani}, {Moran}, {Morello}, {Moses}, {Mukherjee}, {Schlawin}, {Sing}, {Stevenson}, {Taylor}, {Aggarwal}, {Ahrer}, {Allen}, {Barstow}, {Bell}, {Blecic}, {Casewell}, {Chubb}, {Crouzet}, {Cubillos}, {Decin}, {Feinstein}, {Fortney}, {Harrington}, {Heng}, {Iro}, {Kempton}, {Kirk}, {Knutson}, {Krick}, {Leconte}, {Lendl}, {MacDonald}, {Mancini}, {Mansfield}, {May}, {Mayne}, {Miguel}, {Nikolov}, {Ohno}, {Palle}, {Parmentier}, {Petit dit de la Roche}, {Piaulet}, {Powell}, {Rackham}, {Redfield}, {Rogers}, {Rustamkulov}, {Tan}, {Tremblin}, {Tsai}, {Turner}, {de Val-Borro},
  {Venot}, {Welbanks}, {Wheatley}, \& {Zhang}}]{Alderson2023}
{Alderson}, L., {Wakeford}, H.~R., {Alam}, M.~K., {et~al.} 2023, \nat, 614, 664, \dodoi{10.1038/s41586-022-05591-3}

\bibitem[{{Allart} {et~al.}(2020){Allart}, {Pino}, {Lovis}, {Sousa}, {Casasayas-Barris}, {Zapatero Osorio}, {Cretignier}, {Palle}, {Pepe}, {Cristiani}, {Rebolo}, {Santos}, {Borsa}, {Bourrier}, {Demangeon}, {Ehrenreich}, {Lavie}, {Lendl}, {Lillo-Box}, {Micela}, {Oshagh}, {Sozzetti}, {Tabernero}, {Adibekyan}, {Allende Prieto}, {Alibert}, {Amate}, {Benz}, {Bouchy}, {Cabral}, {Dekker}, {D'Odorico}, {Di Marcantonio}, {Dumusque}, {Figueira}, {Genova Santos}, {Gonz{\'a}lez Hern{\'a}ndez}, {Lo Curto}, {Manescau}, {Martins}, {M{\'e}gevand}, {Mehner}, {Molaro}, {Nunes}, {Poretti}, {Riva}, {Su{\'a}rez Mascare{\~n}o}, {Udry}, \& {Zerbi}}]{Allart2020}
{Allart}, R., {Pino}, L., {Lovis}, C., {et~al.} 2020, \aap, 644, A155, \dodoi{10.1051/0004-6361/202039234}

\bibitem[{{Astropy Collaboration} {et~al.}(2013){Astropy Collaboration}, {Robitaille}, {Tollerud}, {Greenfield}, {Droettboom}, {Bray}, {Aldcroft}, {Davis}, {Ginsburg}, {Price-Whelan}, {Kerzendorf}, {Conley}, {Crighton}, {Barbary}, {Muna}, {Ferguson}, {Grollier}, {Parikh}, {Nair}, {Unther}, {Deil}, {Woillez}, {Conseil}, {Kramer}, {Turner}, {Singer}, {Fox}, {Weaver}, {Zabalza}, {Edwards}, {Azalee Bostroem}, {Burke}, {Casey}, {Crawford}, {Dencheva}, {Ely}, {Jenness}, {Labrie}, {Lim}, {Pierfederici}, {Pontzen}, {Ptak}, {Refsdal}, {Servillat}, \& {Streicher}}]{2013A&A...558A..33A}
{Astropy Collaboration}, {Robitaille}, T.~P., {Tollerud}, E.~J., {et~al.} 2013, \aap, 558, A33, \dodoi{10.1051/0004-6361/201322068}

\bibitem[{{Astropy Collaboration} {et~al.}(2018){Astropy Collaboration}, {Price-Whelan}, {Sip{\H{o}}cz}, {G{\"u}nther}, {Lim}, {Crawford}, {Conseil}, {Shupe}, {Craig}, {Dencheva}, {Ginsburg}, {VanderPlas}, {Bradley}, {P{\'e}rez-Su{\'a}rez}, {de Val-Borro}, {Aldcroft}, {Cruz}, {Robitaille}, {Tollerud}, {Ardelean}, {Babej}, {Bach}, {Bachetti}, {Bakanov}, {Bamford}, {Barentsen}, {Barmby}, {Baumbach}, {Berry}, {Biscani}, {Boquien}, {Bostroem}, {Bouma}, {Brammer}, {Bray}, {Breytenbach}, {Buddelmeijer}, {Burke}, {Calderone}, {Cano Rodr{\'\i}guez}, {Cara}, {Cardoso}, {Cheedella}, {Copin}, {Corrales}, {Crichton}, {D'Avella}, {Deil}, {Depagne}, {Dietrich}, {Donath}, {Droettboom}, {Earl}, {Erben}, {Fabbro}, {Ferreira}, {Finethy}, {Fox}, {Garrison}, {Gibbons}, {Goldstein}, {Gommers}, {Greco}, {Greenfield}, {Groener}, {Grollier}, {Hagen}, {Hirst}, {Homeier}, {Horton}, {Hosseinzadeh}, {Hu}, {Hunkeler}, {Ivezi{\'c}}, {Jain}, {Jenness}, {Kanarek}, {Kendrew}, {Kern}, {Kerzendorf}, {Khvalko}, {King}, {Kirkby}, {Kulkarni},
  {Kumar}, {Lee}, {Lenz}, {Littlefair}, {Ma}, {Macleod}, {Mastropietro}, {McCully}, {Montagnac}, {Morris}, {Mueller}, {Mumford}, {Muna}, {Murphy}, {Nelson}, {Nguyen}, {Ninan}, {N{\"o}the}, {Ogaz}, {Oh}, {Parejko}, {Parley}, {Pascual}, {Patil}, {Patil}, {Plunkett}, {Prochaska}, {Rastogi}, {Reddy Janga}, {Sabater}, {Sakurikar}, {Seifert}, {Sherbert}, {Sherwood-Taylor}, {Shih}, {Sick}, {Silbiger}, {Singanamalla}, {Singer}, {Sladen}, {Sooley}, {Sornarajah}, {Streicher}, {Teuben}, {Thomas}, {Tremblay}, {Turner}, {Terr{\'o}n}, {van Kerkwijk}, {de la Vega}, {Watkins}, {Weaver}, {Whitmore}, {Woillez}, {Zabalza}, \& {Astropy Contributors}}]{2018AJ....156..123A}
{Astropy Collaboration}, {Price-Whelan}, A.~M., {Sip{\H{o}}cz}, B.~M., {et~al.} 2018, \aj, 156, 123, \dodoi{10.3847/1538-3881/aabc4f}

\bibitem[{{Batalha} {et~al.}(2020){Batalha}, {Rooney}, \& {Mukherjee}}]{virga_initialrelease}
{Batalha}, N., {Rooney}, C., \& {Mukherjee}, S. 2020, {natashabatalha/virga: Initial Release}, v0.0,  Zenodo, \dodoi{10.5281/zenodo.3759888}

\bibitem[{{Batalha} {et~al.}(2019){Batalha}, {Marley}, {Lewis}, \& {Fortney}}]{PICASO2019}
{Batalha}, N.~E., {Marley}, M.~S., {Lewis}, N.~K., \& {Fortney}, J.~J. 2019, \apj, 878, 70, \dodoi{10.3847/1538-4357/ab1b51}

\bibitem[{{Boucher} {et~al.}(2023){Boucher}, {Lafreni{\'e}re}, {Pelletier}, {Darveau-Bernier}, {Radica}, {Allart}, {Artigau}, {Cook}, {Debras}, {Doyon}, {Gaidos}, {Benneke}, {Cadieux}, {Carmona}, {Cloutier}, {Cort{\'e}s-Zuleta}, {Cowan}, {Delfosse}, {Donati}, {Fouqu{\'e}}, {Forveille}, {Grankin}, {H{\'e}brard}, {Martins}, {Martioli}, {Masson}, \& {Vinatier}}]{Boucher2023}
{Boucher}, A., {Lafreni{\'e}re}, D., {Pelletier}, S., {et~al.} 2023, \mnras, 522, 5062, \dodoi{10.1093/mnras/stad1247}

\bibitem[{{Buchner} {et~al.}(2014){Buchner}, {Georgakakis}, {Nandra}, {Hsu}, {Rangel}, {Brightman}, {Merloni}, {Salvato}, {Donley}, \& {Kocevski}}]{pymultinest}
{Buchner}, J., {Georgakakis}, A., {Nandra}, K., {et~al.} 2014, \aap, 564, A125, \dodoi{10.1051/0004-6361/201322971}

\bibitem[{{Chen} {et~al.}(2018){Chen}, {Pall{\'e}}, {Welbanks}, {Prieto-Arranz}, {Madhusudhan}, {Gandhi}, {Casasayas-Barris}, {Murgas}, {Nortmann}, {Crouzet}, {Parviainen}, \& {Gandolfi}}]{Chen2018}
{Chen}, G., {Pall{\'e}}, E., {Welbanks}, L., {et~al.} 2018, \aap, 616, A145, \dodoi{10.1051/0004-6361/201833033}

\bibitem[{{Evans} {et~al.}(2018{\natexlab{a}}){Evans}, {Sing}, {Goyal}, {Nikolov}, {Marley}, {Zahnle}, {Henry}, {Barstow}, {Alam}, {Sanz-Forcada}, {Kataria}, {Lewis}, {Lavvas}, {Ballester}, {Ben-Jaffel}, {Blumenthal}, {Bourrier}, {Drummond}, {Garc{\'\i}a Mu{\~n}oz}, {L{\'o}pez-Morales}, {Tremblin}, {Ehrenreich}, {Wakeford}, {Buchhave}, {Lecavelier des Etangs}, {H{\'e}brard}, \& {Williamson}}]{Evans+18}
{Evans}, T.~M., {Sing}, D.~K., {Goyal}, J.~M., {et~al.} 2018{\natexlab{a}}, \aj, 156, 283, \dodoi{10.3847/1538-3881/aaebff}

\bibitem[{{Evans} {et~al.}(2018{\natexlab{b}}){Evans}, {Sing}, {Goyal}, {Nikolov}, {Marley}, {Zahnle}, {Henry}, {Barstow}, {Alam}, {Sanz-Forcada}, {Kataria}, {Lewis}, {Lavvas}, {Ballester}, {Ben-Jaffel}, {Blumenthal}, {Bourrier}, {Drummond}, {Garc{\'\i}a Mu{\~n}oz}, {L{\'o}pez-Morales}, {Tremblin}, {Ehrenreich}, {Wakeford}, {Buchhave}, {Lecavelier des Etangs}, {H{\'e}brard}, \& {Williamson}}]{evans2018wasp121b}
---. 2018{\natexlab{b}}, \aj, 156, 283, \dodoi{10.3847/1538-3881/aaebff}

\bibitem[{{Fairman} {et~al.}(2024){Fairman}, {Wakeford}, \& {MacDonald}}]{fairman2024optical}
{Fairman}, C., {Wakeford}, H.~R., \& {MacDonald}, R.~J. 2024, \aj, 167, 240, \dodoi{10.3847/1538-3881/ad3454}

\bibitem[{{Feinstein} {et~al.}(2023){Feinstein}, {Radica}, {Welbanks}, {Murray}, {Ohno}, {Coulombe}, {Espinoza}, {Bean}, {Teske}, {Benneke}, {Line}, {Rustamkulov}, {Saba}, {Tsiaras}, {Barstow}, {Fortney}, {Gao}, {Knutson}, {MacDonald}, {Mikal-Evans}, {Rackham}, {Taylor}, {Parmentier}, {Batalha}, {Berta-Thompson}, {Carter}, {Changeat}, {dos Santos}, {Gibson}, {Goyal}, {Kreidberg}, {L{\'o}pez-Morales}, {Lothringer}, {Miguel}, {Molaverdikhani}, {Moran}, {Morello}, {Mukherjee}, {Sing}, {Stevenson}, {Wakeford}, {Ahrer}, {Alam}, {Alderson}, {Allen}, {Batalha}, {Bell}, {Blecic}, {Brande}, {Caceres}, {Casewell}, {Chubb}, {Crossfield}, {Crouzet}, {Cubillos}, {Decin}, {D{\'e}sert}, {Harrington}, {Heng}, {Henning}, {Iro}, {Kempton}, {Kendrew}, {Kirk}, {Krick}, {Lagage}, {Lendl}, {Mancini}, {Mansfield}, {May}, {Mayne}, {Nikolov}, {Palle}, {Petit dit de la Roche}, {Piaulet}, {Powell}, {Redfield}, {Rogers}, {Roman}, {Roy}, {Nixon}, {Schlawin}, {Tan}, {Tremblin}, {Turner}, {Venot}, {Waalkes}, {Wheatley}, \&
  {Zhang}}]{Feinstein2023}
{Feinstein}, A.~D., {Radica}, M., {Welbanks}, L., {et~al.} 2023, \nat, 614, 670, \dodoi{10.1038/s41586-022-05674-1}

\bibitem[{{Feroz} \& {Hobson}(2008)}]{Feroz2008}
{Feroz}, F., \& {Hobson}, M.~P. 2008, MNRAS, 384, 449, \dodoi{10.1111/j.1365-2966.2007.12353.x}

\bibitem[{{Flagg} {et~al.}(2023){Flagg}, {Turner}, {Deibert}, {Ridden-Harper}, {de Mooij}, {MacDonald}, {Jayawardhana}, {Gibson}, {Langeveld}, \& {Sing}}]{Flagg2023}
{Flagg}, L., {Turner}, J.~D., {Deibert}, E., {et~al.} 2023, \apjl, 953, L19, \dodoi{10.3847/2041-8213/ace529}

\bibitem[{{Foreman-Mackey} {et~al.}(2013){Foreman-Mackey}, {Hogg}, {Lang}, \& {Goodman}}]{Emcee}
{Foreman-Mackey}, D., {Hogg}, D.~W., {Lang}, D., \& {Goodman}, J. 2013, \pasp, 125, 306, \dodoi{10.1086/670067}

\bibitem[{{Freedman} {et~al.}(2008){Freedman}, {Marley}, \& {Lodders}}]{Freedman2008}
{Freedman}, R.~S., {Marley}, M.~S., \& {Lodders}, K. 2008, \apjs, 174, 504, \dodoi{10.1086/521793}

\bibitem[{{Gao} {et~al.}(2020){Gao}, {Thorngren}, {Lee}, {Fortney}, {Morley}, {Wakeford}, {Powell}, {Stevenson}, \& {Zhang}}]{Gao2020}
{Gao}, P., {Thorngren}, D.~P., {Lee}, E. K.~H., {et~al.} 2020, Nature Astronomy, 4, 951, \dodoi{10.1038/s41550-020-1114-3}

\bibitem[{{Gascon} {et~al.}(2024){Gascon}, {Lopez-Morales}, \& {Wakeford}}]{Carlos}
{Gascon}, C., {Lopez-Morales}, M., \& {Wakeford}, H. 2024, in prep

\bibitem[{{Gharib-Nezhad} {et~al.}(2021){Gharib-Nezhad}, {Iyer}, {Line}, {Freedman}, {Marley}, \& {Batalha}}]{Ehsan2021}
{Gharib-Nezhad}, E., {Iyer}, A.~R., {Line}, M.~R., {et~al.} 2021, \apjs, 254, 34, \dodoi{10.3847/1538-4365/abf504}

\bibitem[{{Gibson} {et~al.}(2017){Gibson}, {Nikolov}, {Sing}, {Barstow}, {Evans}, {Kataria}, \& {Wilson}}]{Gibson2017}
{Gibson}, N.~P., {Nikolov}, N., {Sing}, D.~K., {et~al.} 2017, \mnras, 467, 4591, \dodoi{10.1093/mnras/stx353}

\bibitem[{{Grant} \& {Wakeford}(2022)}]{Exotic2022}
{Grant}, D., \& {Wakeford}, H.~R. 2022, {Exo-TiC/ExoTiC-LD: ExoTiC-LD v3.0.0}, v3.0.0, Zenodo,  Zenodo, \dodoi{10.5281/zenodo.7437681}

\bibitem[{{Grant} {et~al.}(2023){Grant}, {Lewis}, {Wakeford}, {Batalha}, {Glidden}, {Goyal}, {Mullens}, {MacDonald}, {May}, {Seager}, {Stevenson}, {Valenti}, {Visscher}, {Alderson}, {Allen}, {Ca{\~n}as}, {Col{\'o}n}, {Clampin}, {Espinoza}, {Gressier}, {Huang}, {Lin}, {Long}, {Louie}, {Pe{\~n}a-Guerrero}, {Ranjan}, {Sotzen}, {Valentine}, {Anderson}, {Balmer}, {Bellini}, {Hoch}, {Kammerer}, {Libralato}, {Mountain}, {Perrin}, {Pueyo}, {Rickman}, {Rebollido}, {Sohn}, {van der Marel}, \& {Watkins}}]{Grant2023}
{Grant}, D., {Lewis}, N.~K., {Wakeford}, H.~R., {et~al.} 2023, \apjl, 956, L32, \dodoi{10.3847/2041-8213/acfc3b10.3847/2041-8213/acfdab}

\bibitem[{{Guillot}(2010)}]{Guillot2010}
{Guillot}, T. 2010, \aap, 520, A27, \dodoi{10.1051/0004-6361/200913396}

\bibitem[{{Horne}(1986)}]{Horne1986}
{Horne}, K. 1986, \pasp, 98, 609, \dodoi{10.1086/131801}

\bibitem[{{JWST Transiting Exoplanet Community Early Release Science Team} {et~al.}(2023){JWST Transiting Exoplanet Community Early Release Science Team}, {Ahrer}, {Alderson}, {Batalha}, {Batalha}, {Bean}, {Beatty}, {Bell}, {Benneke}, {Berta-Thompson}, {Carter}, {Crossfield}, {Espinoza}, {Feinstein}, {Fortney}, {Gibson}, {Goyal}, {Kempton}, {Kirk}, {Kreidberg}, {L{\'o}pez-Morales}, {Line}, {Lothringer}, {Moran}, {Mukherjee}, {Ohno}, {Parmentier}, {Piaulet}, {Rustamkulov}, {Schlawin}, {Sing}, {Stevenson}, {Wakeford}, {Allen}, {Birkmann}, {Brande}, {Crouzet}, {Cubillos}, {Damiano}, {D{\'e}sert}, {Gao}, {Harrington}, {Hu}, {Kendrew}, {Knutson}, {Lagage}, {Leconte}, {Lendl}, {MacDonald}, {May}, {Miguel}, {Molaverdikhani}, {Moses}, {Murray}, {Nehring}, {Nikolov}, {Petit dit de la Roche}, {Radica}, {Roy}, {Stassun}, {Taylor}, {Waalkes}, {Wachiraphan}, {Welbanks}, {Wheatley}, {Aggarwal}, {Alam}, {Banerjee}, {Barstow}, {Blecic}, {Casewell}, {Changeat}, {Chubb}, {Col{\'o}n}, {Coulombe}, {Daylan}, {de Val-Borro},
  {Decin}, {Dos Santos}, {Flagg}, {France}, {Fu}, {Garc{\'\i}a Mu{\~n}oz}, {Gizis}, {Glidden}, {Grant}, {Heng}, {Henning}, {Hong}, {Inglis}, {Iro}, {Kataria}, {Komacek}, {Krick}, {Lee}, {Lewis}, {Lillo-Box}, {Lustig-Yaeger}, {Mancini}, {Mandell}, {Mansfield}, {Marley}, {Mikal-Evans}, {Morello}, {Nixon}, {Ortiz Ceballos}, {Piette}, {Powell}, {Rackham}, {Ramos-Rosado}, {Rauscher}, {Redfield}, {Rogers}, {Roman}, {Roudier}, {Scarsdale}, {Shkolnik}, {Southworth}, {Spake}, {Steinrueck}, {Tan}, {Teske}, {Tremblin}, {Tsai}, {Tucker}, {Turner}, {Valenti}, {Venot}, {Waldmann}, {Wallack}, {Zhang}, \& {Zieba}}]{ERS2023}
{JWST Transiting Exoplanet Community Early Release Science Team}, {Ahrer}, E.-M., {Alderson}, L., {et~al.} 2023, \nat, 614, 649, \dodoi{10.1038/s41586-022-05269-w}

\bibitem[{{Kanumalla} {et~al.}(2024){Kanumalla}, {Line}, {Weiner Mansfield}, {Welbanks}, {Smith}, {Bean}, {Pino}, {Brogi}, \& {Panwar}}]{Kanumalla2024}
{Kanumalla}, K., {Line}, M.~R., {Weiner Mansfield}, M., {et~al.} 2024, arXiv e-prints, arXiv:2406.14072, \dodoi{10.48550/arXiv.2406.14072}

\bibitem[{{Kawashima} \& {Ikoma}(2019)}]{Kawashima&Ikoma19}
{Kawashima}, Y., \& {Ikoma}, M. 2019, \apj, 877, 109, \dodoi{10.3847/1538-4357/ab1b1d}

\bibitem[{{Kempton} \& {Knutson}(2024)}]{KemptonKnutson2024}
{Kempton}, E. M.~R., \& {Knutson}, H.~A. 2024, Reviews in Mineralogy and Geochemistry, 90, 411, \dodoi{10.2138/rmg.2024.90.12}

\bibitem[{{Kreidberg}(2015)}]{Batman}
{Kreidberg}, L. 2015, \pasp, 127, 1161, \dodoi{10.1086/683602}

\bibitem[{{Kurucz}(1993)}]{Kurucz1993}
{Kurucz}, R. 1993, ATLAS9 Stellar Atmosphere Programs and 2 km/s grid. Kurucz CD-ROM No. 13. Cambridge, 13

\bibitem[{{Laginja} \& {Wakeford}(2020)}]{Exoticism}
{Laginja}, I., \& {Wakeford}, H. 2020, The Journal of Open Source Software, 5, 2281, \dodoi{10.21105/joss.02281}

\bibitem[{{Lam} {et~al.}(2017){Lam}, {Faedi}, {Brown}, {Anderson}, {Delrez}, {Gillon}, {H{\'e}brard}, {Lendl}, {Mancini}, {Southworth}, {Smalley}, {Triaud}, {Turner}, {Hay}, {Armstrong}, {Barros}, {Bonomo}, {Bouchy}, {Boumis}, {Collier Cameron}, {Doyle}, {Hellier}, {Henning}, {Jehin}, {King}, {Kirk}, {Louden}, {Maxted}, {McCormac}, {Osborn}, {Palle}, {Pepe}, {Pollacco}, {Prieto-Arranz}, {Queloz}, {Rey}, {S{\'e}gransan}, {Udry}, {Walker}, {West}, \& {Wheatley}}]{Lam2017}
{Lam}, K.~W.~F., {Faedi}, F., {Brown}, D.~J.~A., {et~al.} 2017, \aap, 599, A3, \dodoi{10.1051/0004-6361/201629403}

\bibitem[{{Lavvas} \& {Koskinen}(2017)}]{Lavvas&Koskinen17}
{Lavvas}, P., \& {Koskinen}, T. 2017, \apj, 847, 32, \dodoi{10.3847/1538-4357/aa88ce}

\bibitem[{Li {et~al.}(2015)Li, Gordon, Rothman, Tan, Hu, Kassi, Campargue, \& Medvedev}]{li2015rovibrational}
Li, G., Gordon, I.~E., Rothman, L.~S., {et~al.} 2015, The Astrophysical Journal Supplement Series, 216, 15

\bibitem[{{Lodders}(2010)}]{Lodders2009}
{Lodders}, K. 2010, in Astrophysics and Space Science Proceedings, Vol.~16, Principles and Perspectives in Cosmochemistry, 379, \dodoi{10.1007/978-3-642-10352-0_8}

\bibitem[{{Lothringer} {et~al.}(2022){Lothringer}, {Sing}, {Rustamkulov}, {Wakeford}, {Stevenson}, {Nikolov}, {Lavvas}, {Spake}, \& {Winch}}]{Lothringer2022}
{Lothringer}, J.~D., {Sing}, D.~K., {Rustamkulov}, Z., {et~al.} 2022, \nat, 604, 49, \dodoi{10.1038/s41586-022-04453-2}

\bibitem[{Lupu {et~al.}(2021)Lupu, Freedman, Gharib-Nezhad, Visscher, \& Molliere}]{LupuCK_picaso}
Lupu, R., Freedman, R., Gharib-Nezhad, E., Visscher, C., \& Molliere, P. 2021, Correlated k coefficients for {H2}-{He} atmospheres; 196 spectral windows and 1460 pressure-temperature points,  [object Object], \dodoi{10.5281/ZENODO.7542068}

\bibitem[{{MacDonald}(2023)}]{MacDonald2023}
{MacDonald}, R.~J. 2023, The Journal of Open Source Software, 8, 4873, \dodoi{10.21105/joss.04873}

\bibitem[{{MacDonald} \& {Madhusudhan}(2017)}]{MacDonald2017}
{MacDonald}, R.~J., \& {Madhusudhan}, N. 2017, \mnras, 469, 1979, \dodoi{10.1093/mnras/stx804}

\bibitem[{{MacDonald} \& {Madhusudhan}(2019)}]{macdonaldmadhu2019hatp26b}
---. 2019, \mnras, 486, 1292, \dodoi{10.1093/mnras/stz789}

\bibitem[{{Magic} {et~al.}(2015){Magic}, {Chiavassa}, {Collet}, \& {Asplund}}]{Magic2015}
{Magic}, Z., {Chiavassa}, A., {Collet}, R., \& {Asplund}, M. 2015, \aap, 573, A90, \dodoi{10.1051/0004-6361/201423804}

\bibitem[{{Mandel} \& {Agol}(2002)}]{MandelAgol2002}
{Mandel}, K., \& {Agol}, E. 2002, \apjl, 580, L171, \dodoi{10.1086/345520}

\bibitem[{{Marley} {et~al.}(2021){Marley}, {Saumon}, {Visscher}, {Lupu}, {Freedman}, {Morley}, {Fortney}, {Seay}, {Smith}, {Teal}, \& {Wang}}]{Marley2021}
{Marley}, M.~S., {Saumon}, D., {Visscher}, C., {et~al.} 2021, \apj, 920, 85, \dodoi{10.3847/1538-4357/ac141d}

\bibitem[{{Marsh}(1989)}]{Marsch1989}
{Marsh}, T.~R. 1989, \pasp, 101, 1032, \dodoi{10.1086/132570}

\bibitem[{{McCullough} {et~al.}(2014){McCullough}, {Crouzet}, {Deming}, \& {Madhusudhan}}]{McCullough+14}
{McCullough}, P.~R., {Crouzet}, N., {Deming}, D., \& {Madhusudhan}, N. 2014, \apj, 791, 55, \dodoi{10.1088/0004-637X/791/1/55}

\bibitem[{{Morley} {et~al.}(2012){Morley}, {Fortney}, {Marley}, {Visscher}, {Saumon}, \& {Leggett}}]{Marley2012}
{Morley}, C.~V., {Fortney}, J.~J., {Marley}, M.~S., {et~al.} 2012, \apj, 756, 172, \dodoi{10.1088/0004-637X/756/2/172}

\bibitem[{{Mukherjee} {et~al.}(2023){Mukherjee}, {Batalha}, {Fortney}, \& {Marley}}]{PICASO2023}
{Mukherjee}, S., {Batalha}, N.~E., {Fortney}, J.~J., \& {Marley}, M.~S. 2023, \apj, 942, 71, \dodoi{10.3847/1538-4357/ac9f48}

\bibitem[{{Nortmann} {et~al.}(2024){Nortmann}, {Lesjak}, {Yan}, {Cont}, {Czesla}, {Lavail}, {Rains}, {Nagel}, {Boldt-Christmas}, {Hatzes}, {Reiners}, {Piskunov}, {Kochukhov}, {Heiter}, {Shulyak}, {Rengel}, \& {Seemann}}]{Nortmann2024}
{Nortmann}, L., {Lesjak}, F., {Yan}, F., {et~al.} 2024, arXiv e-prints, arXiv:2404.12363, \dodoi{10.48550/arXiv.2404.12363}

\bibitem[{{Ohno} \& {Kawashima}(2020)}]{Ohno2020}
{Ohno}, K., \& {Kawashima}, Y. 2020, \apjl, 895, L47, \dodoi{10.3847/2041-8213/ab93d7}

\bibitem[{{Ohno} {et~al.}(2020){Ohno}, {Okuzumi}, \& {Tazaki}}]{Ohno+20_fluffy}
{Ohno}, K., {Okuzumi}, S., \& {Tazaki}, R. 2020, \apj, 891, 131, \dodoi{10.3847/1538-4357/ab44bd}

\bibitem[{{Ormel} \& {Min}(2019)}]{Ormel&Min19}
{Ormel}, C.~W., \& {Min}, M. 2019, \aap, 622, A121, \dodoi{10.1051/0004-6361/201833678}

\bibitem[{{Palle} {et~al.}(2017){Palle}, {Chen}, {Prieto-Arranz}, {Nowak}, {Murgas}, {Nortmann}, {Pollacco}, {Lam}, {Montanes-Rodriguez}, {Parviainen}, \& {Casasayas-Barris}}]{Palle2017}
{Palle}, E., {Chen}, G., {Prieto-Arranz}, J., {et~al.} 2017, \aap, 602, L15, \dodoi{10.1051/0004-6361/201731018}

\bibitem[{{Parmentier} {et~al.}(2016){Parmentier}, {Fortney}, {Showman}, {Morley}, \& {Marley}}]{Parm2016}
{Parmentier}, V., {Fortney}, J.~J., {Showman}, A.~P., {Morley}, C., \& {Marley}, M.~S. 2016, \apj, 828, 22, \dodoi{10.3847/0004-637X/828/1/22}

\bibitem[{{Pinhas} \& {Madhusudhan}(2017)}]{Pinhas2017}
{Pinhas}, A., \& {Madhusudhan}, N. 2017, \mnras, 471, 4355, \dodoi{10.1093/mnras/stx1849}

\bibitem[{{Pirzkal}(2020)}]{Pirzkal2020}
{Pirzkal}, N. 2020, {Updated Calibration of the UVIS G280 Grism}, Instrument Science Report WFC3 2020-9, 27 pages

\bibitem[{{Pirzkal} \& {Ryan}(2017)}]{PirzkalRyan2017}
{Pirzkal}, N., \& {Ryan}, R. 2017, {A more generalized coordinate transformation approach for grisms}, Instrument Science Report WFC3 2017-01 (v.1), 9 pages

\bibitem[{Polyansky {et~al.}(2018)Polyansky, Kyuberis, Zobov, Tennyson, Yurchenko, \& Lodi}]{polyansky2018exomol}
Polyansky, O.~L., Kyuberis, A.~A., Zobov, N.~F., {et~al.} 2018, Monthly Notices of the Royal Astronomical Society, 480, 2597

\bibitem[{{Pont} {et~al.}(2013){Pont}, {Sing}, {Gibson}, {Aigrain}, {Henry}, \& {Husnoo}}]{STIS2013}
{Pont}, F., {Sing}, D.~K., {Gibson}, N.~P., {et~al.} 2013, \mnras, 432, 2917, \dodoi{10.1093/mnras/stt651}

\bibitem[{{Powell} {et~al.}(2019){Powell}, {Louden}, {Kreidberg}, {Zhang}, {Gao}, \& {Parmentier}}]{Powell+19}
{Powell}, D., {Louden}, T., {Kreidberg}, L., {et~al.} 2019, \apj, 887, 170, \dodoi{10.3847/1538-4357/ab55d9}

\bibitem[{{Rooney} {et~al.}(2022){Rooney}, {Batalha}, {Gao}, \& {Marley}}]{virga_details}
{Rooney}, C.~M., {Batalha}, N.~E., {Gao}, P., \& {Marley}, M.~S. 2022, \apj, 925, 33, \dodoi{10.3847/1538-4357/ac307a}

\bibitem[{{Rustamkulov} {et~al.}(2023){Rustamkulov}, {Sing}, {Mukherjee}, {May}, {Kirk}, {Schlawin}, {Line}, {Piaulet}, {Carter}, {Batalha}, {Goyal}, {L{\'o}pez-Morales}, {Lothringer}, {MacDonald}, {Moran}, {Stevenson}, {Wakeford}, {Espinoza}, {Bean}, {Batalha}, {Benneke}, {Berta-Thompson}, {Crossfield}, {Gao}, {Kreidberg}, {Powell}, {Cubillos}, {Gibson}, {Leconte}, {Molaverdikhani}, {Nikolov}, {Parmentier}, {Roy}, {Taylor}, {Turner}, {Wheatley}, {Aggarwal}, {Ahrer}, {Alam}, {Alderson}, {Allen}, {Banerjee}, {Barat}, {Barrado}, {Barstow}, {Bell}, {Blecic}, {Brande}, {Casewell}, {Changeat}, {Chubb}, {Crouzet}, {Daylan}, {Decin}, {D{\'e}sert}, {Mikal-Evans}, {Feinstein}, {Flagg}, {Fortney}, {Harrington}, {Heng}, {Hong}, {Hu}, {Iro}, {Kataria}, {Kempton}, {Krick}, {Lendl}, {Lillo-Box}, {Louca}, {Lustig-Yaeger}, {Mancini}, {Mansfield}, {Mayne}, {Miguel}, {Morello}, {Ohno}, {Palle}, {Petit dit de la Roche}, {Rackham}, {Radica}, {Ramos-Rosado}, {Redfield}, {Rogers}, {Shkolnik}, {Southworth}, {Teske}, {Tremblin},
  {Tucker}, {Venot}, {Waalkes}, {Welbanks}, {Zhang}, \& {Zieba}}]{Rustamkulov2023}
{Rustamkulov}, Z., {Sing}, D.~K., {Mukherjee}, S., {et~al.} 2023, \nat, 614, 659, \dodoi{10.1038/s41586-022-05677-y}

\bibitem[{{Ryabchikova} {et~al.}(2015){Ryabchikova}, {Piskunov}, {Kurucz}, {Stempels}, {Heiter}, {Pakhomov}, \& {Barklem}}]{Ryabchikova2015}
{Ryabchikova}, T., {Piskunov}, N., {Kurucz}, R.~L., {et~al.} 2015, Physica Scripta, 90, 054005, \dodoi{10.1088/0031-8949/90/5/054005}

\bibitem[{Sharp \& Burrows(2007)}]{sharp2007atomic}
Sharp, C.~M., \& Burrows, A. 2007, The Astrophysical Journal Supplement Series, 168, 140

\bibitem[{{Sing} {et~al.}(2015){Sing}, {Wakeford}, {Showman}, {Nikolov}, {Fortney}, {Burrows}, {Ballester}, {Deming}, {Aigrain}, {D{\'e}sert}, {Gibson}, {Henry}, {Knutson}, {Lecavelier des Etangs}, {Pont}, {Vidal-Madjar}, {Williamson}, \& {Wilson}}]{STIS2015}
{Sing}, D.~K., {Wakeford}, H.~R., {Showman}, A.~P., {et~al.} 2015, \mnras, 446, 2428, \dodoi{10.1093/mnras/stu2279}

\bibitem[{{Sing} {et~al.}(2019){Sing}, {Lavvas}, {Ballester}, {Lecavelier des Etangs}, {Marley}, {Nikolov}, {Ben-Jaffel}, {Bourrier}, {Buchhave}, {Deming}, {Ehrenreich}, {Mikal-Evans}, {Kataria}, {Lewis}, {L{\'o}pez-Morales}, {Garc{\'\i}a Mu{\~n}oz}, {Henry}, {Sanz-Forcada}, {Spake}, {Wakeford}, \& {PanCET Collaboration}}]{Sing2019}
{Sing}, D.~K., {Lavvas}, P., {Ballester}, G.~E., {et~al.} 2019, \aj, 158, 91, \dodoi{10.3847/1538-3881/ab2986}

\bibitem[{{Skaf} {et~al.}(2020){Skaf}, {Bieger}, {Edwards}, {Changeat}, {Morvan}, {Kiefer}, {Blain}, {Zingales}, {Poveda}, {Al-Refaie}, {Baeyens}, {Gressier}, {Guilluy}, {Jaziri}, {Modirrousta-Galian}, {Mugnai}, {Pluriel}, {Whiteford}, {Wright}, {Yip}, {Charnay}, {Leconte}, {Drossart}, {Tsiaras}, {Venot}, {Waldmann}, \& {Beaulieu}}]{Skaf2020}
{Skaf}, N., {Bieger}, M.~F., {Edwards}, B., {et~al.} 2020, \aj, 160, 109, \dodoi{10.3847/1538-3881/ab94a3}

\bibitem[{{Sotzen} {et~al.}(2020){Sotzen}, {Stevenson}, {Sing}, {Kilpatrick}, {Wakeford}, {Filippazzo}, {Lewis}, {H{\"o}rst}, {L{\'o}pez-Morales}, {Henry}, {Buchhave}, {Ehrenreich}, {Fraine}, {Garc{\'\i}a Mu{\~n}oz}, {Jayaraman}, {Lavvas}, {Lecavelier des Etangs}, {Marley}, {Nikolov}, {Rathcke}, \& {Sanz-Forcada}}]{sotzen2020wasp127b}
{Sotzen}, K.~S., {Stevenson}, K.~B., {Sing}, D.~K., {et~al.} 2020, \aj, 159, 5, \dodoi{10.3847/1538-3881/ab5442}

\bibitem[{{Spake} {et~al.}(2021){Spake}, {Sing}, {Wakeford}, {Nikolov}, {Mikal-Evans}, {Deming}, {Barstow}, {Anderson}, {Carter}, {Gillon}, {Goyal}, {Hebrard}, {Hellier}, {Kataria}, {Lam}, {Triaud}, \& {Wheatley}}]{Spake2021}
{Spake}, J.~J., {Sing}, D.~K., {Wakeford}, H.~R., {et~al.} 2021, \mnras, 500, 4042, \dodoi{10.1093/mnras/staa3116}

\bibitem[{{Steinrueck} {et~al.}(2023){Steinrueck}, {Koskinen}, {Lavvas}, {Parmentier}, {Zieba}, {Tan}, {Zhang}, \& {Kreidberg}}]{Steinrueck+23}
{Steinrueck}, M.~E., {Koskinen}, T., {Lavvas}, P., {et~al.} 2023, \apj, 951, 117, \dodoi{10.3847/1538-4357/acd4bb}

\bibitem[{Tashkun \& Perevalov(2011)}]{tashkun2011cdsd}
Tashkun, S., \& Perevalov, V. 2011, Journal of Quantitative Spectroscopy and Radiative Transfer, 112, 1403

\bibitem[{{Thorngren} {et~al.}(2019){Thorngren}, {Gao}, \& {Fortney}}]{Thorngren2019}
{Thorngren}, D., {Gao}, P., \& {Fortney}, J.~J. 2019, \apjl, 884, L6, \dodoi{10.3847/2041-8213/ab43d0}

\bibitem[{{van Dokkum}(2001)}]{vanDokkum2001}
{van Dokkum}, P.~G. 2001, \pasp, 113, 1420, \dodoi{10.1086/323894}

\bibitem[{Virtanen {et~al.}(2020)Virtanen, Gommers, Oliphant, Haberland, Reddy, Cournapeau, Burovski, Peterson, Weckesser, Bright, {van der Walt}, Brett, Wilson, Millman, Mayorov, Nelson, Jones, Kern, Larson, Carey, Polat, Feng, Moore, {VanderPlas}, Laxalde, Perktold, Cimrman, Henriksen, Quintero, Harris, Archibald, Ribeiro, Pedregosa, {van Mulbregt}, \& {SciPy 1.0 Contributors}}]{Scipy}
Virtanen, P., Gommers, R., Oliphant, T.~E., {et~al.} 2020, Nature Methods, 17, 261, \dodoi{10.1038/s41592-019-0686-2}

\bibitem[{{Visscher} {et~al.}(2010){Visscher}, {Lodders}, \& {Fegley}}]{visscher2010chemistry}
{Visscher}, C., {Lodders}, K., \& {Fegley}, Bruce, J. 2010, \apj, 716, 1060, \dodoi{10.1088/0004-637X/716/2/1060}

\bibitem[{{{\v{Z}}{\'a}k} {et~al.}(2019){{\v{Z}}{\'a}k}, {Kab{\'a}th}, {Boffin}, {Ivanov}, \& {Skarka}}]{Zak2019}
{{\v{Z}}{\'a}k}, J., {Kab{\'a}th}, P., {Boffin}, H. M.~J., {Ivanov}, V.~D., \& {Skarka}, M. 2019, \aj, 158, 120, \dodoi{10.3847/1538-3881/ab32ec}

\bibitem[{{Wakeford} {et~al.}(2022){Wakeford}, {Alderson}, {Batalha}, {Grant}, {Lewis}, {Lopez-Morales}, {MacDonald}, {Marley}, {Moran}, \& {Ohno}}]{HUSTLE2022}
{Wakeford}, H., {Alderson}, L., {Batalha}, N., {et~al.} 2022, {Hubble Ultraviolet-optical Survey of Transiting Legacy Exoplanets (HUSTLE) treasury program}, HST Proposal. Cycle 30, ID. \#17183

\bibitem[{{Wakeford} \& {Sing}(2015)}]{WakefordSing2015}
{Wakeford}, H.~R., \& {Sing}, D.~K. 2015, \aap, 573, A122, \dodoi{10.1051/0004-6361/201424207}

\bibitem[{{Wakeford} {et~al.}(2016){Wakeford}, {Sing}, {Evans}, {Deming}, \& {Mandell}}]{Wakeford2016}
{Wakeford}, H.~R., {Sing}, D.~K., {Evans}, T., {Deming}, D., \& {Mandell}, A. 2016, \apj, 819, 10, \dodoi{10.3847/0004-637X/819/1/10}

\bibitem[{{Wakeford} {et~al.}(2017){Wakeford}, {Visscher}, {Lewis}, {Kataria}, {Marley}, {Fortney}, \& {Mandell}}]{Wakeford2017}
{Wakeford}, H.~R., {Visscher}, C., {Lewis}, N.~K., {et~al.} 2017, \mnras, 464, 4247, \dodoi{10.1093/mnras/stw2639}

\bibitem[{{Wakeford} {et~al.}(2018){Wakeford}, {Sing}, {Deming}, {Lewis}, {Goyal}, {Wilson}, {Barstow}, {Kataria}, {Drummond}, {Evans}, {Carter}, {Nikolov}, {Knutson}, {Ballester}, \& {Mandell}}]{Wakeford2018}
{Wakeford}, H.~R., {Sing}, D.~K., {Deming}, D., {et~al.} 2018, \aj, 155, 29, \dodoi{10.3847/1538-3881/aa9e4e}

\bibitem[{{Wakeford} {et~al.}(2020){Wakeford}, {Sing}, {Stevenson}, {Lewis}, {Pirzkal}, {Wilson}, {Goyal}, {Kataria}, {Mikal-Evans}, {Nikolov}, \& {Spake}}]{Wakeford2020}
{Wakeford}, H.~R., {Sing}, D.~K., {Stevenson}, K.~B., {et~al.} 2020, \aj, 159, 204, \dodoi{10.3847/1538-3881/ab7b78}

\bibitem[{Wende {et~al.}(2010)Wende, Reiners, Seifahrt, \& Bernath}]{wende2010crires}
Wende, S., Reiners, A., Seifahrt, A., \& Bernath, P. 2010, Astronomy \& Astrophysics, 523, A58

\bibitem[{{Wilson}(2021)}]{wilson2021chisqcdf}
{Wilson}, T.~J. 2021, Research Notes of the American Astronomical Society, 5, 265, \dodoi{10.3847/2515-5172/ac3984}

\bibitem[{{Woitke} {et~al.}(2018){Woitke}, {Helling}, {Hunter}, {Millard}, {Turner}, {Worters}, {Blecic}, \& {Stock}}]{woitke2018chemistry}
{Woitke}, P., {Helling}, C., {Hunter}, G.~H., {et~al.} 2018, \aap, 614, A1, \dodoi{10.1051/0004-6361/201732193}

\bibitem[{Yurchenko {et~al.}(2017)Yurchenko, Amundsen, Tennyson, \& Waldmann}]{yurchenko2017hybrid}
Yurchenko, S.~N., Amundsen, D.~S., Tennyson, J., \& Waldmann, I.~P. 2017, Astronomy \& Astrophysics, 605, A95

\end{thebibliography}
\bibliographystyle{aasjournal}



\end{document}